\documentclass[12pt,a4]{article}

\usepackage{amsmath,amsthm,amssymb}
\usepackage{a4wide}
\usepackage[dvips]{graphicx}

\usepackage{mybbold}    

\usepackage{def}

\usepackage{wrapfig}
\usepackage[bf,small]{caption2}

\usepackage{color}

\newcommand{\changed}[1]{{#1}}

\begin{document}

\thispagestyle{empty}

\noindent
June 2005, \changed{revised January 2006}\\

\vspace*{1cm}

\begin{center}

{\LARGE\bf 
Semiclassical limits for the\\[1ex] QCD Dirac operator}\\
\vspace*{3cm}
{\large Thomas Guhr}%
\footnote{Email address: {\tt thomas.guhr@matfys.lth.se}} 
\quad and \quad {\large Stefan Keppeler}%
\footnote{Email address: {\tt stefan.keppeler@matfys.lth.se}}

\vspace*{1cm}
{\it Matematisk Fysik,
Lunds Tekniska H\"ogskola,
Lunds Universitet\\
Box 118, SE-22100 Lund, Sweden}
\end{center}

\vfill\vfill

\begin{abstract}
We identify three semiclassical parameters in the QCD Dirac operator. 
Mutual coupling of the different types of degrees of freedom 
(translational, colour and spin) depends on how the semiclassical limit
is taken. We discuss various semiclassical limits and their potential to 
describe spectrum and spectral statistics of the QCD Dirac operator close 
to zero virtuality. 
\end{abstract}
\vfill

\newpage

\section{Introduction}

Quantum chromo dynamics (QCD) is generally believed to be the correct 
theory for describing the strong force between quarks and gluons. 
The property of asymptotic freedom makes it possible to use perturbation theory
at large momentum transfer and allows for a precise description of 
many scattering experiments as carried out in the big accelerator facilities.

As the fundamental theory for the interaction of quarks and gluons it 
also has to be able to describe bound states of quarks and anti-quarks, 
i.e. hadronic matter such as the proton, the neutron, pions etc. 
However, in this energy regime ordinary perturbation theory (expansion 
in the coupling constant) is bound to fail and QCD becomes extremely hard to
solve. So far there are no promising analytical approaches at hand which would 
allow for a calculation of hadronic masses from QCD. It is the main 
goal of lattice gauge theory to numerically calculate 
hadron masses from first principles. 

Within lattice gauge theory QCD is not formulated in the continuum but 
on a discrete and finite space or space-time lattice. 
Hadronic masses can then be extracted from the decay
of fermionic correlation functions. In Euclidean lattice gauge theory 
these correlation functions are given by Euclidean path integrals which 
in turn can be evaluated numerically by Monte Carlo methods. 
The fermionic degrees of freedom are formally integrated out and the
Monte Carlo integration is carried out for a bosonic path integral
only, which, however, contains the spectral determinant of the Dirac
operator in the integration measure.

The evaluation of fermionic determinants, which now has to be
carried out for each update of the gauge field configuration, is
computationally intensive, in particular for realistic, i.e. small,
quark masses. Thus, in the past many studies have been performed in
the so-called quenched approximation in which the fermionic
determinant is neglected completely. This is equivalent to giving the
quarks infinite mass or setting the number of flavours to
zero. Large scale unquenched lattice calculations have only become
available in recent years, and calculations with realistic quark masses
will only be possible with the next generation of specialised super
computers. 

Therefore, any other way of obtaining independent information on the
spectrum of the Dirac operator, and thus the fermionic determinant, is
of great interest. 

In the early nineties it turned out that chiral random matrix theory
(RMT) describes spectral correlations of the QCD Dirac operator
extremely well \cite{ShuVer93,VerZah93} and can even predict the
microscopic spectral density, i.e. the distribution of small
eigenvalues of the Dirac operator, see \cite{VerWet00} for an
overview. However, the somewhat surprising information that the
spectra of lattice QCD are, up to a certain scale, indistinguishable
from the spectra of random matrices cannot be exploited directly in
order to facilitate lattice calculations. The scale mentioned above is
the equivalent of the Thouless energy in disordered systems. It was
theoretically derived in \cite{JanNowPapZah98,OsbVer98} and
identified in lattice QCD data in
\cite{BerGoeGuhJacMaMeySchWeiWetWil98,GuhMaMeyWil99}.

This situation is reminiscent of the situation in low dimensional
quantum chaos. There short range spectral correlations of individual
quantum systems can be described by RMT if the corresponding classical
system is chaotic. In this context a two-fold role is played by
semiclassical methods, in particular by the Gutzwiller trace formula
\cite{Gut71}. On the one hand they provide an explanation for the
correspondence of classical chaoticity and quantum spectral
correlations being described by RMT. On the other hand they also
predict and describe deviations from RMT in long-range correlations,
linking them to non-universal features of short periodic orbits
\cite{Ber85}. Up to now, such a scale could not be identified in
spectra of the QCD Dirac operator for a frozen, i.e.~fixed,
configuration of the gauge fields. We notice that the equivalent of
the Thouless energy mentioned above is an effect due to the
\changed{fluctuation} of the gauge fields and can thus only be seen after
averaging over all configurations, see the discussion in
\cite{GuhWilWei00,GuhWil01b}.

Inspired by this analogy one may ask: Are semiclassical contributions
the missing ingredient which would make it possible to constructively
use the RMT information when calculating fermionic determinants? As a
first step towards an answer we develop semiclassical approaches to
the Euclidean QCD Dirac operator and in particular discuss qualitative
features of the classical dynamics arising in this context. Notice
that the word ``semiclassical'' in this context always refers to
asymptotic statements about the spectrum of the Dirac operator 
\changed{ -- technically a problem in single particle quantum
  mechanics rather than in quantum field theory --}
which is not the same as loop expansions which are also called
``semiclassical'' in quantum field theory.  Thus, our approach is in a
similar spirit as works relating the spectral analysis of the QCD
Dirac operator to the theory of disordered systems
\cite{JanNowPapZah98,OsbVer98}.
 
This article is organised as follows.
In section \ref{sec:qcd-dirac} we review some basic formulae 
and discuss the semiclassical structure of the QCD Dirac operator.
In section \ref{sec:trace_formula} we 
briefly sketch a strategy for deriving trace formulae which we will follow 
in the subsequent sections. 
The discussion of semiclassical approaches to the Dirac operator in Abelian
gauge fields presented in section \ref{sec:abelian} serves as prerequisite 
for our semiclassical analysis for the QCD Dirac operator which follows in 
section \ref{sec:non-abelian}. The latter contains the main results of 
this article identifying three semiclassical parameters and discussing 
the classical dynamics arising in different (combined) semiclassical 
limits. Some details left open in sections \ref{sec:abelian} and 
\ref{sec:non-abelian} are solved by our study of the squared Dirac operator 
in section \ref{sec:squared}. In section \ref{sec:mean_densities} we discuss 
whether and how our theory can be used for describing features of Dirac 
spectra close to zero virtuality. Section \ref{sec:torus} illustrates 
our theory for an explicit example. We conclude with a discussion of our 
findings and by indicating possible future directions of research 
in section \ref{sec:conclusions}.

\section{The QCD Dirac operator}
\label{sec:qcd-dirac}

The free Euclidean Dirac operator describing 
massless spin $1/2$-particles reads 
\begin{equation}
\label{eq:free_Dirac}
  \op{D}= \frac{\hbar}{\ui} \gamma_\mu \partial_\mu \, .
\end{equation}
We adopt the summation convention over repeated Greek indices from $1$ to
$d$, the number of space-time dimensions. 
The $\gamma$-matrices satisfy
\begin{equation}
\label{eq:gamma_anticommutator}
  \{ \gamma_\mu , \gamma_\nu \} = 2\delta_{\mu\nu} \, .
\end{equation}
Describing massive particles simply amounts to adding $-\ui m$ 
to \eqref{eq:free_Dirac}.
Also note that in the context of lattice gauge theory often 
the anti-Hermitean operator $\ui\op{D}$ is called Dirac operator. 
In dimension $d=4$ we will later explicitly use the 
chiral representation, 
\begin{equation}
\label{eq:chiral_rep}
  \vek{\gamma} = 
  \begin{pmatrix} 0 & -\ui\vek{\sigma} \\ \ui\vek{\sigma} & 0 \end{pmatrix} 
  \, , \quad
  \gamma_4 = \begin{pmatrix} 0 & \eins_2 \\ \eins_2 & 0 \end{pmatrix} \, , 
\end{equation} 
where $\eins_n$ denotes the $n\times n$ unit matrix and $\vek{\sigma}$ is the 
three-vector of Pauli matrices,
\begin{equation}
  \sigma_1 = \begin{pmatrix} 0 & 1 \\ 1 & 0 \end{pmatrix} , \quad
  \sigma_2 = \begin{pmatrix} 0 & -\ui \\ \ui & 0 \end{pmatrix} , \quad
  \sigma_3 = \begin{pmatrix} 1 & 0 \\ 0 & -1 \end{pmatrix} .
\end{equation}
In this representation $\gamma_5 = \gamma_1\gamma_2\gamma_3\gamma_4$ reads 
\begin{equation}
  \gamma_5 = \begin{pmatrix} \eins_2 & 0 \\ 0 & -\eins_2 \end{pmatrix} \, .
\end{equation}

\changed{When now introducing a non-Abelian gauge field we put special
emphasis on the appearance of $\hbar$ and its consequences for
semiclassics. In this way we will identify the most natural asymptotic
treatment from the perspective of semiclassical physics. However, by
introducing fields with an $\hbar$-dependent magnitude alternative
options are also possible and we will remark on those in the
appropriate places. Moreover, we will explain which situations in
standard QCD language correspond to the scenarios discussed.}

A non-Abelian gauge field $A_\mu(x)$ is introduced by minimal coupling, 
\begin{equation}
\label{eq:qcd-Dirac}
  \op{D} = 
  \gamma_\mu \left( \frac{\hbar}{\ui} \partial_\mu - \hbar g A_\mu(x) \right) 
  \, . 
\end{equation}
Notice the appearance of $\hbar$, together with the coupling constant $g$ 
which turns the covariant derivative into 
\begin{equation}
\label{eq:covarinat_derivative_QCD}
  D_\mu = \partial_\mu - \ui g A_\mu(x) \, ,
\end{equation}
$\op{D}=-\ui\hbar\gamma_\mu D_\mu$. 
This is different from the Abelian case, i.e. quantum electro dynamics (QED), 
where the minimal coupling prescription reads
\begin{equation}
  \frac{\hbar}{\ui}\partial_\mu 
  \quad \longmapsto \quad 
  \frac{\hbar}{\ui}\partial_\mu - e A_\mu(x) \, .
\end{equation} 
Here we denote the coupling constant,
i.e. the electric charge of the fermion, by $e$ 
(we set $c=1$) and thus the covariant derivative reads 
\begin{equation}
\label{eq:covarinat_derivative_QED}
  D_\mu = \partial_\mu - \ui \frac{e}{\hbar} A_\mu \, .
\end{equation}
The reason for the different appearance of $\hbar$ in these two cases 
is that non-Abelian fields couple to themselves. More precisely, 
when writing down the QCD-Lagrangean which upon variation yields both,
the Dirac equation and the classical Yang-Mills equations for $A_\mu$, 
the latter 
would contain a self-interaction term which would explicitly depend on 
$\hbar$ if the covariant derivative \eqref{eq:covarinat_derivative_QCD} 
had the same $\hbar$-dependence as \eqref{eq:covarinat_derivative_QED}.
Since this cannot be true for a classical equation, formula
\eqref{eq:covarinat_derivative_QCD} is the correct choice for 
non-Abelian fields. 
To illustrate this consider the field strength tensor deriving from 
\eqref{eq:covarinat_derivative_QCD}, 
\begin{equation}
\label{eq:field_strength}
  F_{\mu\nu} = \partial_\mu A_\nu - \partial_\nu A_\mu -\ui g [A_\mu,A_\nu]
\end{equation} 
which does not contain $\hbar$. 
On the other hand an Abelian field does not couple to itself, 
the last term in \eqref{eq:field_strength} vanishes, 
and therefore in QED a covariant derivative of the form 
\eqref{eq:covarinat_derivative_QED} is allowed, because it 
does not lead to an $\hbar$-dependence of the field strength.
The very same mechanism is responsible for the well-known fact that 
one can have elementary particles with different electric charges 
but that all particles which couple to the non-Abelian colour field 
do so with the same coupling constant, 
i.e. they all have the same colour charge. 

\changed{Notice that the observation described between
\eqref{eq:qcd-Dirac} and here holds true as long as all
$\hbar$-dependencies in the formulae are displayed explicitly,
i.e. quantities such as $g$, $A_\mu$ or $F_{\mu\nu}$ do not depend on
$\hbar$.
In particular, the argumentation as laid out above is independent of the
scaling properties of the QCD-action. If one, e.g., rescales the
fields according to $A_\mu = A_\mu^\prime/\sqrt{\hbar}$ then eqs.
\eqref{eq:qcd-Dirac} and \eqref{eq:field_strength} will read
$\op{D}=\gamma_\mu\left(
\frac{\hbar}{\ui}\partial_\mu-\hbar^{3/2}gA_\mu^\prime \right)$ and
$F_{\mu\nu} = \sqrt{\hbar} \partial_\mu A_\nu^\prime - \sqrt{\hbar}
\partial_\nu A_\mu^\prime - \ui \hbar g [A_\mu^\prime,A_\nu^\prime]$,
respectively.  Thus, we have formally produced powers of $\hbar$ in
unfamiliar places. However, as long as the original fields $A_\mu$ are
of order $1$ then the rescaled fields $A_\mu^\prime$ are of order
$\hbar^{-1/2}$, and therefore the $\hbar$-dependence of the couplings
between fermion and gauge field and of the gauge field to itself are
as before.

The situation changes if we, instead of just rescaling the fields, consider
fields whose order of magnitude is $\hbar$-dependent. For instance, a
gauge field of order $1/\hbar$ gives rise to a Dirac operator in which
the coupling of fermion and colour field has the same
$\hbar$-dependence as in QED. In QCD such a strong field is called an
external colour field. 

In the situation, however, which was described between
\eqref{eq:qcd-Dirac} and \eqref{eq:field_strength}, the
electromagnetic fields are external fields whereas the colour fields
are microscopic or dynamical fields. Since the $\hbar$-dependence of
the latter is chosen such that the classical field equations are
$\hbar$-independent it is also common in QCD to speak of a ``classical
gauge field'' in this context. From the point of view of high energy
physics it may appear slightly inconsistent to discuss external
electro-magnetic fields and microscopic colour fields in the same
context. From the point of view of semiclassical physics, however, we
have treated both types of fields on the same footing.

In the following we will concentrate on the situation with microscopic
gauge fields, i.e. on the Dirac operator \eqref{eq:qcd-Dirac} where
all $\hbar$-dependencies are displayed explicitly. A treatment of
external colour fields would lead to different semiclassical
asymptotics.}

In order to \changed{shed some more light on} the physics behind
\changed{the $\hbar$-dependence discussed above}, decompose the
non-Abelian field in terms of the generators $X^a$ of the gauge group
$G$, say $\SU(N)$,
\begin{equation}
  A_\mu = \frac{1}{2} A_\mu^a X^a \, ,
\end{equation}
where summation over the repeated Latin index $a$ is from $1$ to the 
dimension of the Lie algebra. 
The $X^a$ are traceless, Hermitean $N\times N$-matrices satisfying the 
Lie algebra relations 
\begin{equation}
  [X^a,X^b] = f^{abc} X^c \, ,
\end{equation}
with structure constants $f^{abc}$, and are normalised according to 
\begin{equation}
  \utr(X^aX^b) = 2 \delta_{ab} \, .
\end{equation}
If we now view 
\begin{equation}
\label{eq:color-operator}
  \op{C}^a := \frac{\hbar}{2} X^a
\end{equation}
as the quantum observable describing the colour degrees of freedom 
(of the fermion) the Dirac operator \eqref{eq:qcd-Dirac} takes the form 
\begin{equation}
  \op{D}= \gamma_\mu \left(\op{p}_\mu - g \op{C}^a A_\mu^a(x) \right) \, ,
\end{equation}
with the colour and momentum operators $\op{C}^a$ and $\op{p}_\mu$, 
representing the quantisation of some classical observables 
$C^a$ and $p_\mu$.

The point of view adopted in \eqref{eq:color-operator} is typical for 
internal, i.e. microscopic, degrees of freedom, a familiar example  
being the non-relativistic spin operator $\op{\vecs}=\hbar\vecsig/2$,
which has the same structure as \eqref{eq:color-operator} 
with $X^a$ replaced by the Pauli matrices, the generators of $\SU(2)$. 

For later reference let us also introduce the matrix valued function 
on classical phase space,
\begin{equation}
\label{eq:symbol-qcd-dirac}
  D(p,x) = 
  \gamma_\mu \left( p_\mu - \frac{\hbar g}{2} X^a A_\mu^a(x) \right) \, ,
\end{equation}
from which the QCD Dirac operator can be obtained by replacing $p_\mu$ 
with $-\ui\hbar\partial_\mu$. In microlocal analysis or 
Wigner-Weyl calculus \eqref{eq:symbol-qcd-dirac} 
is referred to as the Weyl symbol of the 
Dirac operator \eqref{eq:qcd-Dirac}, which in turn can be obtained
from its symbol by Weyl quantisation,
\begin{equation}
  (\op{D}\Psi)(x) = \frac{1}{(2\pi\hbar)^d} \int_{\R^d} \int_{\R^d} 
  D\left(p,\frac{x+y}{2}\right) \, \ue^{\frac{\ui}{\hbar} p_\mu(x_\mu-y_\mu)}
  \, \Psi(y) \, \ud^d p \, \ud^d y \, .
\end{equation}
Wigner-Weyl calculus is a particularly useful tool when studying 
semiclassical asymptotics. In a setting where the semiclassical limit 
is identified with $\hbar\to0$ one would classify the terms in 
\eqref{eq:symbol-qcd-dirac} according to their $\hbar$-dependence as
the principal symbol 
\begin{equation}
\label{eq:principal_symbol}
  D_0(p,x) = \gamma_\mu p_\mu
\end{equation}
and the sub-principal symbol 
\begin{equation}
\label{eq:sub-principal_symbol}
  D_1(p,x) = - \frac{g}{2}  \gamma_\mu X^a A_\mu^a(x) \, ,
\end{equation}
respectively. Eventually we will also use the notation 
\begin{equation}
  \mathrm{symb}[\op{D}](p,x) \equiv D(p,x)
\end{equation}
for the Weyl symbol of an operator.

\section{Semiclassical trace formulae}
\label{sec:trace_formula}

Before we go into details about the semiclassics for the QCD Dirac operator
let us say a few words about semiclassical trace formulae in 
general and briefly sketch one method for deriving them; for details, 
however, we refer to the cited literature. 

We are interested in the spectrum of the Hermitean operator $\op{D}$.
For simplicity assume that the spectrum is pure point, i.e. we have a set of
eigenvalues $\lambda_n$ and a complete ortho-normal set of corresponding 
eigenstates $\Psi_n$, 
\begin{equation}
  \op{D} \Psi_n = \lambda_n \Psi_n \, .
\end{equation}
Our main focus lies on the spectral density 
\begin{equation}
  \rho(\lambda) = \sum_n \delta(\lambda-\lambda_n) \, ,
\end{equation}
\changed{which is gauge invariant.}
In order to derive a semiclassical expression for $\rho(\lambda)$ consider
the evolution equation 
\begin{equation}
\label{eq:evolution} 
  \ui \hbar \frac{\partial}{\partial t} \Psi(x,t) = \op{D} \Psi(x,t) \, .
\end{equation}
Note that the time parameter $t$ is not the physical time but an auxiliary 
variable. The physical time is already included in the components of 
$x$ and we are dealing with the Euclidean Dirac operator. 
Likewise the spectral parameter $\lambda$ is not an energy but referred 
to as virtuality. 
Now define \changed{the evolution kernel} $K(x,y,t)$ by 
\begin{equation}
  \Psi(x,t) = \int_{\R^d} K(x,y,t) \, \Psi(y,0) \, \ud^d y \, ,
\end{equation}
which has the spectral representation 
\begin{equation}
  K(x,y,t) = \sum_n \Psi_n(x) \, \Psi_n^\dag(y) \, 
             \ue^{-\frac{\ui}{\hbar}\lambda_n t} \, .
\end{equation}
Obviously, $K(x,y,t)$ also has to solve \eqref{eq:evolution} with 
initial condition
\begin{equation}
  K(x,y,0) = \delta(x-y) \, .
\end{equation}
By Fourier transforming \changed{the evolution kernel} and taking the trace 
on $L^2(\R^d) \otimes \C^{\dimrep} \otimes \C^4$, 
where $\dimrep$ denotes the dimension 
of the representation of the gauge group, we obtain the spectral density,
\begin{equation}
\label{eq:density_aa_trace}
  \utr \changed{\, \frac{1}{2\pi\hbar}} 
       \int_{\R^d} \int_{-\infty}^\infty K(x,x,t) \, 
       \ue^{\frac{\ui}{\hbar}\lambda t} \, \ud t \, \ud^d x 
  = \rho(\lambda) 
\end{equation}
Here $\utr$ denotes the trace over the matrix degrees of freedom.

In oder to obtain a semiclassical approximation for the spectral density,
one can begin with a WKB-type ansatz for the time evolution kernel,
\begin{equation}
\label{eq:sc_ansatz}
  K(x,y,t) = \frac{1}{(2\pi\hbar)^d} \int_{\R^d} 
  [a_0(x,\xi,t) + \hbar a_1(x,\xi,t) + \hdots ] \, 
  \ue^{\frac{\ui}{\hbar}(S(x,\xi,t) - \xi y)} \, \ud^d\xi \, .
\end{equation}
Inserting into \eqref{eq:evolution} and sorting by powers of $\hbar$ one finds 
a sequence of equations which can be solved 
order by order yielding $S,a_0,a_1,\hdots$.
In leading order one always finds a Hamilton-Jacobi equation for the 
phase $S$, 
\begin{equation}
\label{eq:HJG_allgemein}
  \Lambda(\nab{x} S, x) + \frac{\partial S}{\partial t} = 0 \, , 
\end{equation}
with a classical Hamiltonian $\Lambda$ given by an eigenvalue of 
the principal symbol of $\op{D}$. Classical Hamilton-Jacobi theory now 
tells us that the solution $S$ of \eqref{eq:HJG_allgemein} generates 
classical dynamics from the phase space point $(\xi,\nab{\xi}S)$ to 
$(\nab{x}S,x)$ in time $t$, showing that the integration parameter 
$\xi$ of the ansatz \eqref{eq:sc_ansatz} plays the role of an initial 
momentum for the classical system. 

In order to derive a semiclassical approximation to $\rho(\lambda)$,
one also needs to determine the leading order amplitude $a_0$ 
which is fixed by the next-to-leading order equation. The latter,
usually referred to as transport equation, 
has the following structure,
\begin{equation}
\label{eq:simple_transport} 
  \left( \frac{\partial}{\partial t} + (\nab{p} \Lambda)
  \nab{x} \right) a_0 
  + \frac{1}{2} \left( \frac{\partial^2 \Lambda}{\partial x_\mu \partial p_\mu}  + \frac{\partial^2 \Lambda}{\partial p_\mu \partial p_\nu} \frac{\partial^2 S}{\partial x_\mu \partial x_\nu} \right) a_0
  + \hdots = 0
\end{equation}
\changed{The reader easily verifies this structure by explictly doing
  the calculation for an operator of his choice. A derivation of the
  general result can, e.g., be found in appendix E of \cite{Kep03b}.} 

The first \changed{bracket on the l.h.s. of \eqref{eq:simple_transport}}
is a derivative along the trajectory generated by the 
$\Lambda$-dynamics, whereas the second term, roughly speaking, measures 
the behaviour of neighbouring phase space points. Without additional terms 
\eqref{eq:simple_transport} is solved by 
$\sqrt{\det \partial^2S/\partial x \partial \xi}$\changed{, see, e.g.,
appendix B of \cite{Kep03b} for a compact derivation}. 
If, besides the terms displayed explicitly in \eqref{eq:simple_transport}, 
further contributions show up in the transport equation then they 
represent the transport of internal degrees of freedom 
(such as spin or colour as we will see below) along the trajectory of the flow 
with Hamiltonian $\Lambda$. 

Therefore, in order to find the full classical system corresponding to the 
quantum Hamiltonian $\op{D}$ one has to 
(i) determine the Hamiltonian(s) $\Lambda$ and 
(ii) carefully analyse all contributions to the transport equation. 

Having determined $S(x,\xi,t)$ and $a(x,\xi,t)$, 
i.e. having obtained a semiclassical approximation to the kernel $K(x,y,t)$, 
a trace formula can be derived in a straight-forward manner by inserting 
everything into \eqref{eq:density_aa_trace} and evaluating all integrals 
in leading order with the method  of stationary phase.

The stationarity conditions for the $x$- and $\xi$-integrals read
\begin{equation}
  \nab{x} S(x,\xi,t) = \xi
  \quad \text{and} \quad
  \nab{\xi}S(x,\xi,t) = x \, .
\end{equation}
According to classical Hamilton-Jacobi theory this means that both 
initial and final momentum as well as initial and final position of the 
trajectory generated by $S$ have to be identical. Thus, only phase space 
points that lie on periodic orbits contribute to the semiclassical expression
for $\rho(\lambda)$.

A special role is played by the periodic points of period zero which 
are given by the whole hypersurface of constant virtuality $\lambda$,
\begin{equation}
\label{eq:virtuality_shell}
  \Omega_\lambda := \{ (p,x) \ | \ \Lambda(p,x) = \lambda \} \, .
\end{equation}
Since their action is also zero they yield the only non-oscillating 
contribution to the spectral density and thus constitute the mean density,
often called Weyl term, 
\begin{equation}
\label{eq:Weyl_general}
  \overline{\rho}(\lambda) 
  = \frac{|\Omega_\lambda|}{(2\pi\hbar)^d}
  = \int_{\R^d} \int_{\R^d} \delta(\Lambda(p,x)-\lambda) \, 
    \frac{\ud^d p \, \ud^d x}{(2\pi\hbar)^d} \, , 
\end{equation}
with a possible multiplicity factor deriving from the internal 
degrees of freedom such as spin or colour. 

Finally we obtain the following general structure for a 
semiclassical trace formula,
\begin{equation}
\label{eq:general_trace_formula}
  \rho(\lambda) \sim \overline{\rho}(\lambda) 
  + \sum_\gamma \mathcal{A}_\gamma(\lambda) \, 
                \ue^{\frac{\ui}{\hbar}S_\gamma(\lambda)} \, .
\end{equation}
Here $\gamma$ labels both, isolated periodic orbits and 
larger families of periodic points, like, e.g., Liouville-Arnold 
tori in integrable systems. The amplitudes $\mathcal{A}_\gamma(\lambda)$
are derived by keeping track of all contributions in the various 
stationary phase approximations involved. 

If one is interested in the precise mathematical meaning of this 
distributional identity and in an absolutely convergent version of 
the trace formula, which can be used for numerical calculations, 
it is convenient to multiply the expressions with a test function in 
$t$ before taking the Fourier transforms, 
see e.g. \cite{Mei92, PauUri95,BolKep99a}

\section{Semiclassical parameters in the Abelian case}
\label{sec:abelian}

In this section we discuss semiclassical approximations to the 
Dirac operator in Abelian gauge fields. We will keep the presentation short 
and closely follow similar studies for the Dirac Hamiltonian which were 
carried out in \cite{BolKep98,BolKep99a}, however, pointing out small 
differences which are due to the fact that we are dealing with the 
Euclidean Dirac operator instead. The results obtained here 
will also be needed for our discussion of the non-Abelian case in the 
following section. 

Consider the equation of motion \eqref{eq:evolution} 
for the time evolution kernel, 
\begin{equation}
\label{eq:eom_kernel}
  \ui\hbar \frac{\partial}{\partial t} \, K(x,y,t) = \op{D} \, K(x,y,t) \, , 
\end{equation}
where the derivatives in the Dirac Hamiltonian,
\begin{equation}
  \op{D} = 
  \gamma_\mu \left( \frac{\hbar}{\ui} \partial_\mu - e A_\mu(x) \right) \, ,
\end{equation}
with Abelian $A_\mu$, act on the first argument of $K$. 
Inserting an ansatz of type \eqref{eq:sc_ansatz} with 
scalar phase $S$ and matrix-valued amplitudes $a_k$ 
into the evolution equation we find in leading orders
\begin{align}
\label{eq:hbar^0_abel}
  \left[ \frac{\partial S}{\partial t} + D(\nab{x}S,x) \right] a_0 &= 0 \, ,
  \\
\label{eq:hbar^1_abel}
  \left[ \frac{\partial S}{\partial t} + D(\nab{x}S,x) \right] a_1
  + \left( \frac{\partial}{\partial t} + \gamma_\mu \partial_\mu \right) a_0 
  &= 0 \, , 
\end{align}
where 
\begin{equation}
D(p,x) = 
\begin{pmatrix} 0 & \pi_4 - \ui\vecsig \vecpi\\
               \pi_4 + \ui\vecsig \vecpi & 0 \end{pmatrix}
\end{equation}
is the (principal) symbol of $\op{D}$ and 
$\pi_\mu := p_\mu - e A_\mu(x)$ denotes the kinetic momenta. 
For \eqref{eq:hbar^0_abel} to have non-trivial solutions the term in 
square brackets must have an eigenvalue zero. The eigenvalues of $D(p,x)$ 
are given by 
\begin{equation}
  \Lambda^\pm(p,x) = \pm \sqrt{\pi_\mu \pi_\mu} =: \pm \Lambda \, , 
\end{equation}
both having multiplicity two. We collect the corresponding eigenvectors 
columnwise in the $2\times4$-matrices
\begin{equation}
  V_+(p,x) = \frac{1}{\sqrt{2}} \begin{pmatrix} \eins_2 \\ 
             \frac{\pi_4 + \ui \vecsig\vecpi}{\Lambda} \end{pmatrix} 
  \, , \quad 
  V_-(p,x) = \frac{1}{\sqrt{2}} \begin{pmatrix} 
  \frac{\pi_4 - \ui \vecsig\vecpi}{\Lambda} \\ -\eins_2\end{pmatrix} \, .
\end{equation} 
Thus, the solvability condition for \eqref{eq:hbar^0_abel} yields the 
Hamilton-Jacobi equations 
\begin{equation}
\label{eq:HJG_abelian}
  \Lambda^\pm(\nab{x} S^\pm, x) + \frac{\partial S^\pm}{\partial t} = 0 \, ,
\end{equation}
and the general solution of \eqref{eq:eom_kernel} is a superposition 
of terms with positive $(+)$ and negative $(-)$ virtuality. 

Equation \eqref{eq:HJG_abelian} alone does not solve \eqref{eq:hbar^0_abel},
but in addition the leading order amplitude has to satisfy 
$D a_0^\pm = \Lambda^\pm a_0^\pm$. This is guaranteed by the following ansatz,
\begin{equation}
  a_0^\pm(x,\xi,t) 
  = V_\pm(\nab{x}S^\pm,x) \, b_\pm(x,\xi,t) \, V_\pm^\dag(\xi,y)
\end{equation}
where the $2\times2$-matrices $b_\pm$ have to be determined by 
\eqref{eq:hbar^1_abel}. To this end we multiply \eqref{eq:hbar^1_abel}
with $V_\pm^\dag(\nab{x}S^\pm,x)$ from the left and $V_\pm(\xi,y)$ 
from the right, yielding
\begin{equation}
\label{eq:projected_transport}
  V_\pm^\dag(\nab{x}S^\pm,x)
  \left( \gamma_\mu \partial_\mu + \frac{\partial}{\partial t} \right)
  V_\pm(\nab{x}S^\pm,x) \,  b_\pm = 0 \, ,
\end{equation}
\changed{since 
\begin{equation}
  V_\pm^\dag(\nab{x}S^\pm,x) 
  \left[ \frac{\partial S}{\partial t} + D(\nab{x}S,x) \right] = 0 \, . 
\end{equation}
}After a lengthy calculation, which is sketched in 
appendix \ref{app:transport}, one finds
\begin{align}
\label{eq:total_time_derivative}
  V_\pm^\dag V_\pm \frac{\partial b_\pm}{\partial t} 
  + V_\pm^\dag \gamma_\mu V_\pm (\partial_\mu b_\pm) 
  &= \left( \frac{\partial}{\partial t} 
    + \frac{\partial\Lambda^\pm}{\partial p_\mu} \, \partial_\mu \right) b_\pm 
  =: \dot{b}_\pm \, , \\
\label{eq:transport_terms}
  V_\pm^\dag \frac{\partial}{\partial t} V_\pm
  + V_\pm^\dag \gamma_\mu \partial_\mu V_\pm^\dag
  &= \frac{1}{2} 
  \left( \frac{\partial^2 \Lambda^\pm}{\partial x_\mu \partial p_\mu}  
    + \frac{\partial^2 \Lambda^\pm}{\partial p_\mu \partial p_\nu} 
      \frac{\partial^2 S^\pm}{\partial x_\mu \partial x_\nu} \right) 
  - \frac{\ui e}{2\Lambda^\pm} \vecsig (\vecE \pm \vecB) \, , 
\end{align}
where the dot in \eqref{eq:total_time_derivative} denotes a derivative 
along the trajectory generated by $S^\pm$. In addition we have introduced 
the electric and magnetic components, $\vecE$ and $\vecB$, of the field 
strength $F_{\mu\nu}$, according to 
\begin{equation}
\label{eq:electric/magnetic}
F = \begin{pmatrix} 
       0 &  B_3 & -B_2 & -E_1\\
    -B_3 &    0 &  B_1 & -E_2\\
     B_2 & -B_1 &    0 & -E_3\\
     E_1 &  E_2 &  E_3 &    0 
    \end{pmatrix} \, .
\end{equation}
Since we already know how to solve a transport equation of type 
\eqref{eq:simple_transport} the product ansatz
\begin{equation}
\label{eq:product_ansatz}
  b_\pm = \sqrt{\det \frac{\partial^2 S^\pm}{\partial x \partial \xi}} \, d_\pm
\end{equation}
with a $2\times2$ matrix $d$ lends itself to simplify the transport 
equation to 
\begin{equation}
\label{eq:spin-transport}
  \dot{d}_\pm - \frac{\ui e}{2\Lambda^\pm} \vecsig (\vecE \pm \vecB) \, d_\pm
  = 0 \, .
\end{equation}
This equation describes the transport of the spin degrees of freedom 
along the trajectory determined by the Hamiltonian $\Lambda^\pm$. 
Obviously $d_\pm$ takes values in $\SU(2)$ and in the trace formula 
the contribution of each periodic orbit is weighted with the trace of the
corresponding $d_\pm$, i.e. with a character.

Equation~\eqref{eq:spin-transport} can be mapped from $\SU(2)$ to $S^2$ 
by looking at the time evolution of the expectation value $\vecs$ 
of the spin operator \changed{$\frac{\hbar}{2} \vecsig$ in an
arbitrary state $u\in\C^2$ 
-- i.e. $\vecs_\pm = u^\dag d_\pm^\dag \frac{\hbar}{2} \vecsig d_\pm u$ --}
as induced by \eqref{eq:spin-transport},
\begin{equation}
\label{eq:spin-precession}
  \dot{\vecs}_\pm = \vecs_\pm \times \frac{e}{\Lambda^\pm} 
  (\vecE \pm \vecB) \, .
\end{equation}
This equation describes classical spin precession, i.e. it is a Euclidean 
analogue of the Thomas- or Bargman-Michel-Telegdi(BMT)-equation 
\cite{Tho27,BarMicTel59}.
Although \eqref{eq:spin-precession} looks like an equation for the 
three-vector $\vecs$ classical spin precession actually takes place on
the sphere $S^2$ since total spin, i.e. $|\vecs|^2$, is conserved. 
The two-sphere in turn is a symplectic manifold and 
\eqref{eq:spin-precession} defines a volume-preserving flow on it. 
These facts together justify the notion of ``classical spin dynamics'' 
in this context. 
Mathematically speaking, we map the equation from (the representation of) 
the group to its coadjoint orbit, see e.g. \cite{Kir76,BolGla04}. 

Had we dealt with a particle with higher spin from the beginning we would
have obtained a similar spin transport equation as \eqref{eq:spin-transport} 
with only $\vecsig$ replaced by generators of a higher dimensional 
representation of $\SU(2)$ 
and $d_\pm$ now taking values in that representation. 
The weight factor \changed{$\utr d_\pm$} in the trace formula would still be 
a character and the analogous mapping to the sphere would lead to exactly 
the same classical spin precession equation. 
The character entering the trace formula is always completely determined 
by classical spin precession \cite{Kep03b}.

We have thus identified the total classical dynamics arising from a 
semiclassical analysis of the Euclidean Dirac operator as a combination 
of the Hamiltonian flows with Hamiltonians $\Lambda^\pm(p,x)$ accompanied by
spin precession along the orbits. 
Since there is no back-reaction of spin dynamics on the Hamiltonian part 
the total dynamics can be formulated as a skew product flow, either on
$\R^{2d}\times\SU(2)$ or $\R^{2d}\times S^2$ \cite{BolKep99b,BolGlaKep01}.

So far we have discussed what happens in the single semiclassical limit 
$\hbar\to0$. However, also the limit of large spin can be considered as
a semiclassical limit, cf. the so-called kicked top \cite{Haa01}. 
If one simultaneously lets $\hbar\to0$ and 
$s\to\infty$, where $2s+1$ denotes the dimension of the representation 
of $\SU(2)$, such that the product $\hbar s$ is kept non-zero and finite,
also the back-reaction of spin on the translational degrees of freedom 
shows up in the classical picture, 
see e.g. \cite{PleAmaMehBra02,PleZai03,BolGla05}.
We emphasise that, although claimed otherwise in \cite{PleZai03}, 
even for Hamiltonians linear in the spin degrees of freedom
semiclassical asymptotics can only then display both, spin dynamics 
and back reaction, simultaneously 
\changed{in leading semiclassical order}
if one considers the large spin limit. 
This says, however, nothing about the possible 
practical use of this type of approximation even when the actual value of 
$s$ is rather small.
\changed{The situation here is somehow reminiscent of the limit of
  large colour, $N \to \infty$, \cite{tHo74}, which yields valuable
  insights although we are mostly interested in $N=3$.}

We can almost write down the Hamiltonians $\Lambda^\pm(p,x,\vecs)$
for the combined dynamics already with the information gathered so far. 
Omitting the spin-dependent terms it has to reduce to $\Lambda^\pm(p,x)$, 
i.e., formally, $\Lambda^\pm(p,x,0) = \Lambda^\pm(p,x)$, and 
it has to give rise to the spin precession \eqref{eq:spin-precession}.
The relativistic Pauli Hamiltonian, 
\begin{equation}
\label{eq:relativistic_Pauli}
  \Lambda_{\rm Pauli}^\pm(p,x,\vecs) 
  = \Lambda^\pm(p,x) - \frac{e}{\Lambda^\pm(p,x)} \vecs (\vecE(x)\pm\vecB(x)) 
  \, , 
\end{equation}
fulfils these requirements, but so does, e.g., the alternative 
square root type Hamiltonian
\begin{equation}
\label{eq:square_root_Hamiltonian}
  \Lambda_{\rm sqrt}^\pm(p,x,\vecs) = 
  \pm \sqrt{ (p_\mu-eA_\mu(x))(p_\mu-eA_\mu(x)) - 2e \vecs (\vecE(x)\pm\vecB(x)) }
\end{equation}
Both types of Hamiltonians agree in the limit of small spin contribution 
(for illustration one may formally consider the limit $\vecs\to0$) 
but in general they 
lead to different back reactions of spin on the translational degrees of 
freedom. This difference becomes particularly important for small
virtualities $\lambda$.
In section \ref{sec:squared} we will dicuss a simple method for deciding 
which Hamiltonians to use, without explicitly developing a full symbol 
calculus for the combined limits.

\section{Semiclassical parameters in the non-Abelian case}
\label{sec:non-abelian}

With a discussion of semiclassical parameters and limits 
of the QCD Dirac operator and the different classical dynamics arising 
in this context the present section contains the central results of this work.
We perform our analysis along the same lines as laid out in sections 
\ref{sec:trace_formula} and \ref{sec:abelian} and build on the results 
obtained in section \ref{sec:abelian}.

Consider the Dirac operator \eqref{eq:qcd-Dirac} with Weyl symbol 
\eqref{eq:symbol-qcd-dirac}. The time evolution kernel $K(x,y,t)$ 
is now a $4J\times4J$ matrix, where $J$ denotes the dimension of the 
representation of the gauge group, i.e. $J=3$ for QCD with $\SU(3)$
gauge fields in the fundamental representation. Inserting an ansatz of 
type \eqref{eq:sc_ansatz} into the equation of motion \eqref{eq:eom_kernel}
with Dirac operator \eqref{eq:qcd-Dirac} yields in leading orders
\begin{align}
\label{eq:hbar^0_non-abel}
  \left[ \frac{\partial S}{\partial t} + D_0(\nab{x}S,x) \right] a_0 &= 0 \, ,
  \\
\label{eq:hbar^1_non-abel}
  \left[ \frac{\partial S}{\partial t} + D_0(\nab{x}S,x) \right] a_1
  + \left( \frac{\partial}{\partial t} + \gamma_\mu \partial_\mu 
           + D_1(\nab{x}S,x) \right) a_0 
  &= 0 \, , 
\end{align}
where we have used the notation for the principal and sub-principal symbol 
which was introduced in eqs.~\eqref{eq:principal_symbol} and 
\eqref{eq:sub-principal_symbol}. 
The principal symbol has eigenvalues
\begin{equation}
\label{eq:free_classical_Hamiltonian}
  \Lambda^\pm(p,x) = \pm \sqrt{p_\mu p_\mu} =: \pm \Lambda
\end{equation}
with corresponding eigenvectors \changed{collected in the matrices}
\begin{equation}
\label{eq:eigenvectors_non-abel}
  V_+ = \frac{1}{\sqrt{2}} 
        \begin{pmatrix} \eins_2 \\ 
        \frac{p_0 + \ui \vecsig\vecp}{\Lambda} \end{pmatrix} \, , \quad 
  V_- = \frac{1}{\sqrt{2}} 
  \begin{pmatrix} \frac{p_0 - \ui \vecsig\vecp}{\Lambda} \\ 
                  -\eins_2\end{pmatrix} \, .
\end{equation} 
As in section \ref{sec:abelian}, eq.~\eqref{eq:hbar^0_non-abel} demands
that the phase of the ansatz \eqref{eq:sc_ansatz} solves a Hamilton-Jacobi
equation, 
\begin{equation}
  \Lambda^\pm(\nab{x}S^\pm,x) + \frac{\partial S^\pm}{\partial t} = 0 \, , 
\end{equation}
and suggests the following ansatz for the leading order amplitude, 
\begin{equation}
  a_0^\pm(x,\xi,t) 
  = V_\pm(\nab{x}S^\pm,x) \, b_\pm(x,\xi,t) \, V_\pm^\dag(\xi,y) \, .
\end{equation}
The projected transport equation for the $2J\times2J$ matrix $b_\pm$ reads
\begin{equation}
  V_\pm^\dag
  \left( \gamma_\mu \partial_\mu + \frac{\partial}{\partial t} 
  - \ui \frac{g}{2} \gamma_\mu A_\mu^a X^a \right)
  V_\pm b_\pm = 0 \, ,
\end{equation}  
which, using eqs.~\eqref{eq:total_time_derivative} and 
\eqref{eq:transport_terms} with the substitution $\pi_\mu \to p_\mu$, 
cf. \eqref{eq:eigenvectors_non-abel}, simplifies to,
\begin{equation}
  \dot{b}_\pm 
  + \frac{1}{2} 
  \left( \frac{\partial^2 \Lambda^\pm}{\partial x_\mu \partial p_\mu}  
    + \frac{\partial^2 \Lambda^\pm}{\partial p_\mu \partial p_\nu} 
      \frac{\partial^2 S^\pm}{\partial x_\mu \partial x_\nu} \right) \, b_\pm
  - \ui \frac{g}{2} \frac{\partial\Lambda^\pm}{\partial p_\mu} 
    A_\mu^a X^a \, b_\pm = 0 \, .
\end{equation}
As in the Abelian case we separate the translational part according to 
\eqref{eq:product_ansatz} and obtain the following equation for the 
$2J\times2J$ matrix $d_\pm$,
\begin{equation}
\label{eq:colour_transport}
  \dot{d}_\pm - \ui \frac{g}{2} \frac{\partial\Lambda^\pm}{\partial p_\mu} 
  A_\mu^a X^a \, d_\pm = 0 \, .
\end{equation}
In contrast to the Abelian case this equation does not involve the 
spin but the colour degrees of freedom. Accordingly we will refer to it as 
colour transport equation.  

As in the case of spin transport we obtain classical dynamics 
by looking at the equation of motion satisfied by the expectation value
$C^a$ of $\hat{C}^a = \frac{\hbar}{2}X^a$,
\begin{equation}
\label{eq:colour_precession}
  \dot{C}^a = - \frac{g}{2} \frac{\partial\Lambda^\pm}{\partial p_\mu}
              f^{abc} A_\mu^b C^c \, ,
\end{equation}
which we call colour precession. \changed{Equation \eqref{eq:colour_precession}
is the colour part of the Wong equations \cite{Won70} to which we will
come back later.}

As in the case of spin precession discussed in the preceeding section 
there are certain conditions restricting the possible values which 
the variables $C^a$ can assume, thus confining the dynamics 
\eqref{eq:colour_precession} to a compact manifold: 
The (representations of the) Casimir operators 
of the gauge algebra are constants of motion for \eqref{eq:colour_transport}
and from those derive constants of motion for the precession equation 
\eqref{eq:colour_precession}. In the case of $\SU(2)$ there is only the 
quadratic Casimir operator (total spin in the preceeding section) 
which confines the precession to the sphere $S^2=\SU(2)/U(1)$.
If the gauge group is $\SU(3)$ then we have two Casimir operators,
one quadratic and one cubic in the generators 
(or components $C^a$ of classical colour). 
The dynamics of the $C^a$ is thus reduced from $\R^8$ ($8$ generators) 
to a six dimensional manifold, 
the flag manifold $\F^3 = \frac{\SU(3)}{\U(1)\times\U(1)}$, 
see e.g. \cite{BoyPerSan01}. 
For the gauge group $\SU(N)$ we would have $\F^N = \SU(N)/\U(1)^{n-1}$, 
instead. 
In all cases these are maximal coadjoint orbits \cite{Kir76}, 
which are not only 
even dimensional but naturally endowed with a symplectic structure, 
thus constituting the classical phase space for internal degrees of 
freedom such as spin or colour. 

Having understood \eqref{eq:colour_transport} as transport equation for 
the colour degrees of freedom and characterised the underlying classical 
phase space and dynamics we can now ask ourselves why the spin degrees of 
freedom do not show up at this level of the semiclassical treatment, 
neither in the Hamiltonians \eqref{eq:free_classical_Hamiltonian} 
nor in the transport equation. The answer is that spin and translational 
degrees of freedom are coupled by the gauge fields and thus only via the 
internal colour degrees of freedom. With both, spin and colour, being 
internal degrees of freedom a coupling between them, which has to involve 
the product of $\hat{\vecs}=\frac{\hbar}{2}\,\vecsig$ and 
$\hat{C}^a=\frac{\hbar}{2}\, X^a$, is automatically at least of order 
$\hbar^2$. Therefore, it does not enter the leading order phases and 
amplitudes of $\hbar\to0$ asymptotics. 

Comparing with the results of the preceeding section we should expect 
a spin precession equation like \eqref{eq:spin-precession} with the 
electric and magnetic fields replaced by their non-Abelian analogues. 
At this point we can thus guess the following semiclassical hierarchy
(which we will confirm to be correct in the following section):
In pure $\hbar\to0$ asymptotics the phase of semiclassical approximations 
is determined by free translational dynamics alone. The leading order 
amplitude is affected by the colour transport along particle orbits, 
whereas there is no back-reaction of colour onto the translational 
degrees of freedom. Spin shows up only as an $\hbar$-correction to the 
amplitude. While spin precession is driven by both translational 
and colour dynamics it does not act back on either of them. 
Thus, we have a double skew product structure. 

Back reaction can be forced to show up explicitly in the semiclassical 
approximations by considering combined limits. To this end choose a 
$J$ dimensional unitary irreducible representation of the gauge group 
and consider the combined limits $\hbar\to0$ and $J\to\infty$ with 
the product $\hbar J$ kept constant. This will lead to colour entering 
on the same level as the translational degrees of freedom, 
i.e. we will have to deal with the minimally coupled classical Hamiltonians
\begin{equation}
\label{eq:Hamiltonians_pxC}
  \Lambda^\pm(p,x,C) = \pm\sqrt{(p_\mu-gA_\mu^a(x)C^a)(p_\mu-gA_\mu^b(x)C^b)} 
  \, .
\end{equation} 
The coupled classical dynamics arising from these Hamiltonians are
known as the Wong equations \cite{Won70,Mon84}.
Spin will enter on the level of the transport equation as for pure 
$\hbar\to0$ asymptotics in the Abelian case. Thus we have moved from 
a double skew product structure to an ordinary skew product. 
\changed{We mention in passing that by performing the additional limit
$J\to\infty$ with $\hbar J$ fixed we have changed the order of
magnitude of the term $g\hbar A_\mu$, appearing in the Dirac operator
\eqref{eq:qcd-Dirac}, from $\hbar$ to $1$. Thus, this situation may
be physically related to that of an external colour field in the
language of QCD, cf. the discussion following \eqref{eq:field_strength}.
Mathematically, however, the scenario introduced here is different.}

If we go even one step further by taking the triple limit
$\hbar\to0$, $J\to\infty$ and $s\to\infty$ 
with the products $\hbar J$ and $\hbar s$ fixed we will find fully coupled 
Hamiltonian dynamics on the total phase space $\R^{2d} \times \F^3 \times S^2$
(for $\SU(3)$-gauge fields). 
As for the relevant Hamiltonians we have to solve the same problem as at the 
end of section \ref{sec:abelian}. With the knowledge obtained so far it could
be either of Pauli type \eqref{eq:relativistic_Pauli} or of 
square root type \eqref{eq:square_root_Hamiltonian},
with $A_\mu$, $\vecE$ and $\vecB$ replaced by their non-Abelian analogues.

\section{The squared Dirac operator:\\ Confirming the semiclassical hierarchy}
\label{sec:squared}

The following study of the squared Dirac operator serves two purposes. 
On the one hand we prove that the semiclassical hierarchy conjectured 
in the preceeding section is correct and on the other hand we determine
the functional form of the classical Hamiltonians corresponding to the
Dirac operator in simultaneous semiclassical limits. 

We calculate the square of the Dirac operator \eqref{eq:qcd-Dirac} and 
determine the Weyl symbol of $\hat{D}^2$ for different symbol calculi.
The principle underlying this approach is that in a symbol calculus there 
exists always a so-called Moyal product which expresses the symbol 
of the product of two operators as an asymptotic expansion in the 
semiclassical parameter(s) in terms of the symbols of the individual operators.
The leading term in this expansion, i.e. the principle symbol of the product,
is always given by the product of the principal symbols. 
Thus, from the appearance (or absence) of certain dynamical variables 
at given order in the symbol of the squared Dirac operator 
we can conclude at which order these variables appear 
in the symbol of the operator itself. 
Moreover, the principal symbol of the squared operator allows us 
to draw conclusions about the functional form of (the eigenvalues of) 
the principal symbol of the operator itself. 

The square of $\hat{D}$ is most conveniently calculated by decomposing 
the products $\gamma_\mu \gamma_\nu$ and $D_\mu D_\nu$ into their 
symmetric and antisymmetric parts, 
\begin{equation}
\begin{split} 
  \op{D}^2 
  &= -\hbar^2 \gamma_\mu \gamma_\nu D_\mu D_\nu
   = -\frac{\hbar^2}{4} 
     \left( \{ \gamma_\mu, \gamma_\nu \} + [ \gamma_\mu, \gamma_\nu ] \right)
     \left( \{ D_\mu, D_\nu \} + [ D_\mu, D_\nu ] \right)\\
  &= -\frac{\hbar^2}{4} 
     \left( \{ \gamma_\mu, \gamma_\nu \} \{ D_\mu, D_\nu \}
            + [ \gamma_\mu, \gamma_\nu ] [ D_\mu, D_\nu ] \right)\\
  &= -\hbar^2 \left( D_\mu D_\mu 
     - \ui \frac{g}{4} [\gamma_\mu,\gamma_\nu] F_{\mu\nu} \right) \, ,
\end{split}
\end{equation}
where on the last line we have used \eqref{eq:gamma_anticommutator} 
and the definition \eqref{eq:field_strength} of the non-Abelian 
field strength tensor, $F_{\mu\nu} =\frac{\ui}{g}[D_\mu,D_\nu]$.
With the representation \eqref{eq:chiral_rep} we have the commutators
\begin{equation}
  [\gamma_4,\vecgamma] 
  = 2\ui \begin{pmatrix} \vecsig & 0 \\ 0 & -\vecsig \end{pmatrix} 
  \quad \text{and} \quad
  [\gamma_j,\gamma_k] = 2\ui \varepsilon_{jkl} 
  \begin{pmatrix} \sigma_l & 0 \\ 0 & \sigma_l \end{pmatrix} 
  \, , \quad
  j,k,l = 1,2,3 \, .
\end{equation}
Introducing colour-electric and colour-magnetic fields as in 
\eqref{eq:electric/magnetic} we finally obtain
\begin{equation}
  \op{D}^2 = -\hbar^2 D_\mu D_\nu 
  - \hbar^2 g \begin{pmatrix} \vecsig (\vecB+\vecE) & 0 \\ 
                              0 & \vecsig (\vecB-\vecE) \end{pmatrix} \, .
\end{equation}
Thus, the matrix-valued Weyl-symbol of $\op{D}^2$ reads 
\begin{equation}
\label{eq:symbol_D^2}
  \mathrm{symb}[\op{D}^2](p,x)  
  = (p_\mu - \hbar g A_\mu)(p_\mu - \hbar g A_\mu)
  - \hbar^2 g \begin{pmatrix} \vecsig (\vecB+\vecE) & 0 \\ 
                              0 & \vecsig (\vecB-\vecE) \end{pmatrix} \, .
\end{equation}
From this we can easily read off at which order in $\hbar$ the different 
degrees of freedom will enter a semiclassical approximation. To this end 
recall that $A_\mu = \frac{1}{2} A_\mu^a X^a$, $\vecE=\frac{1}{2}\vecE^a X^a$ 
and $\vecB=\frac{1}{2}\vecB^a X^a$. At order $\hbar^0$ only the 
translational degrees of freedom show up in $\mathrm{symb}[\op{D}^2]$. 
The colour degrees of freedom, $X^a$, appear for the first time at order 
$\hbar^1$, whereas the spin degrees of freedom are absent unless we proceed 
up to order $\hbar^2$. The first two observations are in agreement with the 
semiclassical analysis of the non-Abelian Dirac operator, and the last one 
provides the missing element in order to prove the semiclassical hierarchy 
\changed{anticipated} at the end of section \ref{sec:non-abelian}.

If, instead of Wigner-Weyl calculus for the translational degrees of freedom 
only, we \changed{used} a symbol calculus which also maps 
the internal matrix degrees of freedom, spin and colour, to
classical variables, i.e. $\mathrm{symb}[\frac{\hbar}{2}\vecsig]=\vecs$ and
$\mathrm{symb}[\frac{\hbar}{2}X^a]=C^a$, then the symbol of $\hat{D}^2$
reads
\begin{equation}
\label{eq:symbol_D^2_triple}
\mathrm{symb}[\op{D}^2](p,x)  
  = (p_\mu - g A_\mu^a C^a)(p_\mu - g A_\mu^b C^b)
  - 2 g \begin{pmatrix} \vecs (\vecB^a+\vecE^a) C^a & 0 \\ 
                              0 & \vecs (\vecB^a-\vecE^a) C^a\end{pmatrix} \, .
\end{equation}
In the simultaneous limit $\hbar\to0$, $J\to\infty$ and $s\to\infty$ 
with $\hbar J$ and $\hbar s$ fixed this total symbol consists of a 
(diagonal $2\times2$) principal symbol only, i.e. there are no higher order 
terms in any of the three semiclassical parameters. 
The linearity of \eqref{eq:symbol_D^2_triple} in the spin degrees of freedom 
shows that the eigenvalues of the symbol of the Dirac operator $\hat{D}$ 
itself have to be of square root type \eqref{eq:relativistic_Pauli} 
rather than of Pauli type \eqref{eq:square_root_Hamiltonian}.

We remark that in the Abelian case \eqref{eq:symbol_D^2} reads 
\begin{equation}
  \mathrm{symb}[\op{D}^2](p,x)  
  = (p_\mu - e A_\mu)(p_\mu - e A_\mu)
  - \hbar e \begin{pmatrix} \vecsig (\vecB+\vecE) & 0 \\ 
                            0 & \vecsig (\vecB-\vecE) \end{pmatrix} \, ,
\end{equation}
which is consistent with spin appearing in the
leading order transport equation in pure $\hbar\to0$ asymptotics.

\section{Mean density in stochastic fields}
\label{sec:mean_densities}

We have motivated this study with the question whether and how 
semiclassics can be of use for the understanding of spectral 
properties of the QCD Dirac operator. Of interest are here in 
particular spectral functions averaged over an ensemble of gauge 
fields, as they appear in the calculation of path integrals in 
(lattice) quantum field theory. An important example is
the averaged spectral density,
\begin{equation}
  \langle \rho(\lambda) \rangle 
  := \int \rho(\lambda) \, \ue^{S[A]} \, \mathcal{D}A \, ,
\end{equation}
where the action $S[A]$, and thus the integration measure, 
can, e.g., be just the Yang-Mills action (quenched approximation)
or the bosonic part of the full QCD action, including fermionic determinants. 

Prominent features of $\langle \rho(\lambda) \rangle$ are the so-called 
chiral condensate, a non-zero value at virtuality $\lambda=0$, 
and a universal functional form for small values of the virtuality.  
Due to the Banks-Casher relation \cite{BanCas80}, 
see also \cite{SmiSte93,VerWet00,Zya00}, $\langle \rho(0) \rangle$
is proportional to the expectation value $\langle \bar{\psi}\psi \rangle$
of the quantised quark fields $\psi$ in the ground state. Thus, a non-zero 
value indicates the spontaneous breaking of chiral symmetry. 
Moreover, after suitably rescaling the virtuality with the chiral condensate,
the microscopic density becomes universal and can be calculated in 
chiral RMT. Notice that the chiral condensate is not determined by the 
number of exact zero modes, which is a topological invariant of the Dirac 
operator, but it arises, in a suitable limit, from the accumulation of 
small but non-zero eigenvalues. 

In this section we investigate whether and to what extent the different 
semiclassical approaches characterised in section \ref{sec:non-abelian}
are able to explain the formation of a chiral condensate on the level of the 
Weyl term \eqref{eq:Weyl_general}, which is the semiclassical description 
of the mean spectral density. For these considerations we restrict the  
$x$-integration to a subset $\mathcal{V}\subset\R^d$ with finite 
volume $V$ and since the density is symmetric about zero it is sufficient 
to consider only positive $\lambda$. 

\subsection{$\hbar \to 0$}

In pure $\hbar\to0$ asymptotics the periodic orbit structure of 
the trace formula and the hypersurface $\Omega_\lambda$ 
\eqref{eq:virtuality_shell} determining the Weyl term are derived 
from the translational degrees of freedom only. Moreover the 
translational dynamics are extremely simple, namely free. Spin and colour 
enter only as multiplicity pre-factors, 
\begin{equation}
\label{eq:Weyl_hbar->0}
\begin{split} 
  \overline{\rho}(\lambda) 
  &= \frac{J (2s+1)}{(2\pi\hbar)^d} \int_\mathcal{V} \int_{\R^d} 
     \delta(\sqrt{p_\mu p_\mu} - \lambda) \, \ud^d p \, \ud^d x\\
  &= \frac{J (2s+1)}{(2\pi\hbar)^d} V \lambda^{d-1} 
     \int_{S^{d-1}} \ud^{d-1}\omega \\
  &= \frac{2 \pi^{d/2} J (2s+1) V}{(2\pi\hbar)^d \, \Gamma(\frac{d}{2})} 
     \, \lambda^{d-1} \, .
\end{split}
\end{equation}
Thus, the Weyl term reduces to the free mean spectral density and 
is in particular independent of the gauge field configuration. 
For $\hbar=1$, dimension $d=4$, $J=N=3$, 
the number of colours, and spin $s=\frac{1}{2}$ the mean density reads
\begin{equation}
\label{eq:Weyl_free_4D}
  \overline{\rho}(\lambda) = \frac{3V}{4\pi^2} \, \lambda^3 \, .
\end{equation}

\subsection{$\hbar \to 0$, $J \to \infty$}

If we take this combined limit then colour precession 
\eqref{eq:colour_precession} appears on the same level as translational 
dynamics. The Hamiltonians
are not only functions of $p$ and $x$ but also of the classical colour
degrees of freedom $C$. In order to be able to integrate over this larger
phase space we need a parameterisation of the colour part. 
If the gauge group is $\SU(N)$ (in a faithful representation) then 
$C$ has $N^2-1$ components. Colour dynamics, however, lives on the 
$N(N-1)$-dimensional manifold $\F^N$. Let $C(\xi)$ be a parametrisation 
of $\F^N$, then the correctly normalised integration measure, which 
complements $\ud^d p \, \ud^d x/(2\pi\hbar)^d$ stemming from 
the translational degrees of freedom, is
\begin{equation}
  \frac{J}{|\F^N|} \ud^{N(N-1)} \xi
\end{equation}
with $|\F^N| = \int_{\F^N} \ud^{N(N-1)} \xi$ being the volume 
of the flag manifold. Thus the Weyl term now reads
\begin{equation} 
\label{eq:Weyl_J->infty}
\overline{\rho}(\lambda) 
  = \frac{J}{|\F^N|} \frac{(2s+1)}{(2\pi\hbar)^d} 
    \int_{\F^N} \int_\mathcal{V} \int_{\R^d} \delta(\Lambda^+(p,x,C(\xi)) - \lambda) 
    \, \ud^d p \, \ud^d x \, \ud^{N(N-1)}\xi \, .
\end{equation}
However, with 
$\Lambda^+=\sqrt{(p_\mu-gA_\mu^a(x)C^a)(p_\mu-gA_\mu^b(x)C^b)}$, 
see \eqref{eq:Hamiltonians_pxC}, 
after a simple shift of variables in the $p$-integrals 
for fixed $x$ and $C$, $p_\mu \mapsto p_\mu+gA_\mu^a(x)C^a$, 
integration over the colour degrees of freedom becomes trivial and 
\eqref{eq:Weyl_J->infty} reduces to \eqref{eq:Weyl_hbar->0} and
thus once more to the free result. 

\subsection{$\hbar \to 0$, $J \to \infty$, $s \to \infty$}
\label{subsec:Weyl_triple_limit}

In this triple limit all degrees of freedom 
-- translational, colour and spin -- appear on the same footing in the
Hamiltonian, 
\begin{equation}
  \Lambda^+=\sqrt{(p_\mu-gA_\mu^a(x)C^a)(p_\mu-gA_\mu^b(x)C^b) 
                  -2g\vecs(\vecE^a(x)+\vecB^a(x))C^a} \, .
\end{equation}
In order to calculate the Weyl term we also need to parametrise the 
phase space of spin, $S^2$, which we do in spherical coordinates denoting 
the solid angle by $\omega$. The correctly normalised measure is 
$\frac{2s+1}{|S^2|} \, \ud^2\omega$, with volume $|S^2|=4\pi$. 
The mean density thus reads
\begin{equation}
  \overline{\rho}(\lambda) = 
  \frac{(2s+1)J}{4\pi |\F^N| (2\pi\hbar)^d} \int_{S^2} \int_{\F^N} 
  \int_\mathcal{V} \int_{\R^d} 
  \delta(\Lambda^+(p,x,C(\xi),\vecs(\omega)) - \lambda) \, 
  \ud^d p \, \ud^d x \, \ud^{N(N-1)}\xi \, \ud^2\omega \, .
\end{equation}
As before we can shift the integration variable in the $p$-integrals 
in order to remove the explicit appearance of the gauge potentials
$A_\mu^a$. 
However, through the field strengths $\vecE$ and $\vecB$ the expression
still depends on the gauge fields and the integration over the internal 
degrees of freedom does not become trivial.
Yet we are able to calculate the mean density if we consider the average 
over an ensemble of gauge fields, which is the function we are interested 
in anyway,
\begin{equation}
  \langle \overline{\rho}(\lambda) \rangle = 
  \int \overline{\rho}(\lambda) \, \mathcal{D}A \, .
\end{equation}
Using the following property, 
\begin{equation}
  \delta(\Lambda^+ - \lambda) = 2\lambda \, \delta({\Lambda^+}^2 - \lambda^2)
  \, ,
\end{equation}
and employing the Fourier representation of the $\delta$-function we have to 
calculate
\begin{equation}
  \langle \overline{\rho}(\lambda) \rangle = 
  \frac{(2s+1)J \lambda}{4\pi^2 |\F^N| (2\pi\hbar)^d} \int 
  \int_\R \int_{S^2} \int_{\F^N} \int_\mathcal{V} \int_{\R^d} 
  \ue^{\ui ({\Lambda^+}^2 - \lambda^2) t}
  \ud^d p \, \ud^d x \, \ud^{N(N-1)}\xi \, \ud^2\omega \, \ud t \, \mathcal{D}A
  \, .
\end{equation}
For simplicity we calculate this expression using stochastic fields. 
More precisely, we assume \changed{locally independent Gaussian
fluctuations} with the same variance $\sigma$ for all components of
$\vecE$ and $\vecB$. \changed{According to a relation derived in
appendix \ref{app:random_measures}, for Weyl terms this is equivalent
to averaging over constant random fields, i.e.} 
\begin{equation}
  \int \hdots \mathcal{D}A 
  \mapsto 
  \frac{1}{(2\pi\sigma^2)^{3N(N-1)}} \int_{\R^{6N(N-1)}} \hdots
  \ue^{-(\vecE^a\vecE^a + \vecB^a \vecB^a)/(2\sigma^2)} \,  
  \ud^{3N(N-1)}E \, \ud^{3N(N-1)}B \, .
\end{equation}
Now the total exponent is quadratic in $\vecE$ and $\vecB$ and an
average over the fields yields, 
\begin{equation}
\begin{split} 
  &\frac{1}{(2\pi\sigma^2)^{3N(N-1)}} \int_{\R^{6N(N-1)}} 
  \ue^{-\ui 2g \vecs(\vecE^a+\vecB^a)C^a t} \, 
  \ue^{-(\vecE^a\vecE^a + \vecB^a \vecB^a)/(2\sigma^2)} \,  
  \ud^{3N(N-1)}E \, \ud^{3N(N-1)}B\\[1ex]
  &\quad = \exp\left( - 4 \sigma^2 g^2 \vecs^2 C^aC^a t^2 \right) \, .
\end{split}
\end{equation}
Since $\vecs^2$ and $C^aC^a$ correspond to the quadratic Casimir operators 
of $\SU(2)$ and $\SU(N)$, respectively, they are constants, i.e. they 
depend only on $s$ and $J$ but not on $\omega$ and $\xi$. Hence, 
\begin{equation}
\label{eq:Weyl_s->infty}
\begin{split} 
  \langle \overline{\rho}(\lambda) \rangle &= 
  \frac{VJ(2s+1)}{\pi(2\pi\hbar)^d} \, \lambda \int_{\R^d} \int_\R
  \ue^{\ui(p_\mu p_\mu - \lambda^2)t - 4\sigma^2 g^2 \vecs^2 C^aC^a t^2} \, 
  \ud t \, \ud^d p\\
  &= \frac{2VJ(2s+1)}{(2\pi\hbar)^d \sqrt{2\pi v^2}} \, \lambda 
     \int_{\R^d} \ue^{-\frac{1}{2v^2}(p_\mu p_\mu - \lambda^2)^2} \, \ud^d p\\
  &= \frac{4\pi^{d/2}VJ(2s+1)}{(2\pi\hbar)^d \, \Gamma(\frac{d}{2})} \, 
     \frac{\lambda}{\sqrt{2\pi v^2}} 
     \int_0^\infty\ue^{-\frac{1}{2v^2}(p^2 - \lambda^2)^2} \, p^{d-1} \, \ud p
  \, ,
\end{split}
\end{equation}
where we have introduced the abbreviation 
$v:= \sqrt{8} \, \sigma g |\vecs| \sqrt{C^aC^a}$. 
This parameter, being proportional to the variance $\sigma$ and the 
coupling constant $g$, is a measure for the strength of the fields. 
The field free situation corresponds to $v=0$ and one easily confirms 
that in this case \eqref{eq:Weyl_s->infty} reduces to
\eqref{eq:Weyl_hbar->0}. 
\changed{Similarly, for large $\lambda$ the
integral expression grows proportional to $\lambda^{d-2}$ and we once more
obtain the free density \eqref{eq:Weyl_hbar->0}. For arbitrary $\lambda$}
the remaining integral in \eqref{eq:Weyl_s->infty} can be expressed 
in terms of generalised Laguerre functions. 

The most important observation is that for $\lambda\to0$ the 
integral converges to a constant,
\begin{equation}
  \frac{1}{\sqrt{2\pi v^2}} 
  \int_0^\infty\ue^{-\frac{p^4}{2v^2}} \, p^{d-1} \, \ud p
  = \frac{(2v^2)^{d/4-1/2}}{4\sqrt{\pi}} \, \Gamma(\tfrac{d}{4}) \, .
\end{equation}
Thus, for small virtualities the mean density now grows linearly
instead of being proportional to $\lambda^{d-1}$  
as in the previous cases. For instance, for $d=4$, where the 
behaviour changes from cubic to linear, this means a dramatic increase 
in the number of small eigenvalues. 
\changed{If the triple limit discussed here was related
to the scenario of a strong external field then the linear density for
small $\lambda$ could be interpreted as the density within the first
Landau band, cf. \cite{ShuSmi97}.
In any case, the behaviour of the density resulting from our
semiclassical approach shows some remarkable features which we discuss
in the following.}

\subsection{Discussion of Weyl terms}
\label{subsec:Weyl_Discussion}

We have calculated the (averaged) Weyl terms for the spectral density 
of the QCD Dirac operator in all 3 different semiclassical limits 
introduced in section \ref{sec:non-abelian}. In two cases the resulting
leading order mean density is just the mean density for the free Dirac 
operator. 
Only in the triple limit, $\hbar\to0$, $J\to\infty$ and $s\to\infty$, 
have we observed a dependence on gauge fields and internal degrees of freedom.
In particular we have derived an increase in the number of small 
eigenvalues.

In this last case we have replaced the QCD or Yang-Mills action in 
the path integral over the colour fields by a Gaussian measure, thus 
neglecting details of the gauge dynamics. The success of random matrix
models and in particular related work on stochastic field theories
\cite{GuhWilWei00,GuhWil01b} makes us believe that our results are 
nevertheless relevant for QCD.


\changed{To further study the accumulation of small eigenvalues as
borne out by \eqref{eq:Weyl_s->infty} we have to examine the integral
expression
\begin{equation}
  \Phi_d(\lambda) := \frac{1}{\sqrt{2\pi v^2}} 
  \int_0^\infty\ue^{-\frac{1}{2v^2}(p^2-\lambda^2)^2} \, p^{d-1} \, \ud p \, .
\end{equation}
Its value in dimension $d=4$ is given by 
\begin{equation}
\begin{split} 
  \Phi_4(\lambda) &= \frac{1}{\sqrt{2\pi v^2}} 
  \int_0^\infty\ue^{-\frac{1}{2v^2}(p^2 - \lambda^2)^2} \, p^{3} \, \ud p \\
  &= \frac{\sqrt{2v^2}}{4\sqrt{\pi}} \ue^{-\frac{\lambda^4}{2v^2}} 
    + \frac{\lambda^2}{4} 
    \left( 1 + \erf\left( \frac{\lambda^2}{\sqrt{2v^2}}\right) \right) \, .
\end{split}
\end{equation}
}For large $\lambda$ the integral is dominated by the last term, i.e.
it grows like  $\lambda^2/2$, which restores
the $\lambda^3$-behaviour of the free mean density \eqref{eq:Weyl_free_4D}.
On the other hand for small $\lambda$ the integral is determined by the 
first term, giving rise to a linear spectral density, i.e. 
(for $d=4$, $J=3$, $\changed{s}=\frac{1}{2}$ and $\hbar=1$)
\begin{equation}
  \langle \overline{\rho}(\lambda) \rangle
  \approx \frac{3V}{(2\pi)^{5/2}} \, \lambda v \, \ue^{-\frac{\lambda^4}{2v^2}}
  \, .
\end{equation} 
\begin{wrapfigure}{R}{7cm}
\hfill\parbox{7cm}{\color{black}
\includegraphics[width=5cm,angle=-90]{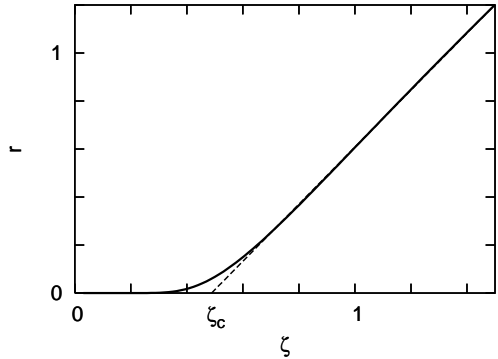}
\caption{\label{fig:chiral?}
Graph of the scaled density $r(\zeta)= \zeta \exp(-1/2\zeta^2)$, 
cf. eqs. \eqref{eq:r_def} and \eqref{eq:zeta_def}, 
which we speculate to be a measure for the chiral condensate.}}
\end{wrapfigure}
It is interesting to see how this expression changes under the variation 
of external parameters. If one wants to consider finite temperatures 
in a field theoretical setting one has to choose an asymmetric
subset $\mathcal{V}\subset\R^4$, say a box with lengths $L_\mu$ 
with fixed $L_4 \ll L_1,L_2,L_3$. Then the inverse of $L_4$ is 
essentially the temperature, see e.g. \cite{MonMue97}.
Having one smaller dimension sets a natural scale for the small 
eigenvalues, namely $\lambda \approx 2\pi/L_4$ (cf. the eigenvalues of the 
free Dirac operator in a box with lengths $L_4 \ll L_1,L_2,L_3$). 
\changed{Since it is this accumulation of small but non-zero
eigenvalues which we want to investigate further, it is instructive to
look at the averaged spectral density on this scale,
\begin{equation}
  \left\langle \overline{\rho}\left(\frac{2\pi}{L_4}\right)\right\rangle
  \approx \frac{3V}{(2\pi)^{3/2}} \, \frac{v}{L_4} 
    \, \exp\left(-\frac{(2\pi)^4}{2v^2L_4^4}\right) \, .
\end{equation}
In fig. \ref{fig:chiral?} we plot the scaled density 
\begin{equation}
\label{eq:r_def}
  r = \frac{L_4^3}{3V\sqrt{2\pi}} \, \langle\overline{\rho}\rangle
\end{equation}
as a function of the scaled variance
\begin{equation}
\label{eq:zeta_def}
  \zeta = \frac{v L_4^2}{(2\pi)^2} \, , 
\end{equation}
observing a curve which is reminiscent of a critical phenomenon with
the spectral density itself playing the r\^ole of the order parameter.
The behaviour for $\zeta > \zeta_c$ would be interpreted as an
indication for a non-zero density for small $\lambda$, which, on the
other hand, vanishes (exponentially) for $\zeta < \zeta_c$.  Thus, we
are tempted to view $\zeta_c$ as a value indicating a phase
transition.  In terms of the original quantities, the variance $v$,
measuring the (coupling) strength of the gauge fields, and the inverse
temperature $L_4$, this implies the following.  For fixed temperature,
on the one hand, the phase transition would occur at a critical
strength of the gauge fields with a vanishing order parameter for weak
fields. For fixed $v$, on the other hand, we would observe the phase
transition for a critical temperature with a non-vanishing order
parameter at low temperatures only. 

We find it remarkable that our simple semiclassical argument is
capable of showing a behaviour which seems to hint at a critical
phenomenon. In view of this we are tempted to put forward the
following speculation.  According to the Bank-Casher relation
\cite{BanCas80} a non-zero averaged density at $\lambda=0$ results
from the formation of a chiral condensate. However, in the derivation
of the Banks-Casher relation the limit $\lambda\to0$ may only be
considered after one has first taken the infinite volume limit
$V\to\infty$, for a suitably normalised expression, and then the
chiral limit $m\to0$ of vanishing sea quark mass(es). If one
interchanges the latter two limits, i.e. if one performs $m\to0$
before $V\to\infty$ then the chiral condensate vanishes, see
e.g. \cite{LeuSmi92}.  In our semiclassical calculation we
do not have a mass parameter, which we could vary accordingly. 

Nevertheless, we find an accumulation of small eigenvalues hinting at
a critical phenomenon. Could it be that this relates to the formation
of a chiral condensate? -- However,} we do not want to conceal that
our discussion only takes into account the leading order Weyl term.
Higher order semiclassical corrections to the mean density may
contribute where the leading order term vanishes, as we will
demonstrate for an example in section \ref{sec:torus}, and the above
discussion also ignores the periodic orbit contributions to the
spectral density.  Recall that a trace formula provides a
decomposition of the density of states $\rho$ into a mean term
$\overline{\rho}$ and a periodic orbit sum $\rho_\mathrm{osc}$,
cf. \eqref{eq:general_trace_formula}, of which only the latter
oscillates as a function of the virtuality $\lambda$. It is now
natural to expect that after averaging over the gauge fields only the
non-oscillating Weyl-term contributes to $\langle \rho \rangle$. The
above discussion is based on this tacit assumption. In general we have
\begin{equation}
  \langle \rho(\lambda) \rangle 
  \sim \langle \overline{\rho}(\lambda) \rangle 
  + \langle \rho_\mathrm{osc}(\lambda) \rangle
\end{equation}
and it is not guaranteed that 
$\langle \varrho_\mathrm{osc} \rangle = 0$ holds 
in a short virtuality interval close to $\lambda = 0$.

If besides the formation of the chiral condensate one also wants 
to explain the universal microscopic density, characteristic of 
the chiral ensembles of random matrix theory and observed in lattice
calculations, cf. \cite{VerWet00}, a theory involving (correlations in) 
the periodic orbit contributions will be required. The analysis in 
\cite{GnuSeiOppZir03,GnuSei04a,GnuSei04b}, 
where the emergence of universal microscopic 
densities is discussed within a graph model should be viewed as a 
guideline which, combined with our semiclassical approximations, would 
put the semiclassical understanding of spectral correlations in QCD 
on a similar level as in (low dimensional) quantum chaos.

\section{Example: Fermions in $\SU(2)$-fields on $\T^2$}
\label{sec:torus}

As an illustration for the structure of semiclassical trace formulae 
and for the calculation of some of the contributions we discuss 
the example of the Dirac operator on a two-dimensional torus $\T^2$
with constant $\SU(2)$ gauge fields. With ``constant field'' we actually 
mean constant potentials $A_\mu^a$, which, due to the non-Abelian 
character, can give rise to a non-vanishing field strength, 
cf. \eqref{eq:field_strength}. 
For this scenario we can analytically calculate the eigenvalues 
and derive an exact trace formula. The contributions to this trace
formula are then compared to the corresponding semiclassical expressions.
In order to keep the presentation simple we only discuss 
the case with fixed representations for the internal degrees of 
freedom, i.e. pure $\hbar\to0$ asymptotics.

Consider a two-dimensional Euclidean Dirac operator, 
in external $\SU(2)$ fields, 
\begin{equation}
\label{eq:2D_Dirac}
  \op{D} = 
  \left( \frac{\hbar}{\ui} \partial_\mu 
         - \frac{\hbar}{2} g \vecA_\mu \vecsig \right) \gamma_\mu \, .
\end{equation}
In two dimensions the $\gamma$-matrices can be chosen of type $2\times2$, e.g
\begin{equation}
  \gamma_1 = \begin{pmatrix} 0 & 1 \\ 1 & 0 \end{pmatrix}
  \, , \quad 
  \gamma_2 = \begin{pmatrix} 0 & -\ui \\ \ui & 0 \end{pmatrix} \, .
\end{equation}
More precisely, we should write 
\begin{equation}
  \gamma_\mu = \eins_2 \otimes \sigma^\mu
\end{equation}
and replace $\vecsig$ in \eqref{eq:2D_Dirac} by $\vecsig \otimes \eins_2$.
As configuration space we choose a two dimensional box with lengths $L_\mu$ 
and periodic boundary conditions,
\begin{equation}
  \Psi(x_1+L_1,x_2) = 
    \Psi(x_1,x_2) \, , \quad
  \Psi(x_1,x_2+L_2) = 
    \Psi(x_1,x_2) \, ,
\end{equation}
i.e. we put the system on a torus $\T^2$. 
With the ansatz $\Psi(x) = u \, \exp(\frac{\ui}{\hbar} p_\mu x_\mu)$ the 
boundary conditions require 
\begin{equation}
\label{eq:quantised_p}
  p_\mu = \frac{2\pi\hbar}{L_\mu} \, n_\mu 
  \, , \quad n_\mu \in \Z
  \qquad \text{(no summation convention!)}
\end{equation}
and the Dirac operator reduces to an ordinary $4\times4$ matrix, 
which has to be diagonalised. The eigenvalues are most conveniently 
determined via the square $\op{D}^2$ of the Dirac operator. By a 
calculation similar to that in section \ref{sec:squared} one finds
\begin{equation}
\label{eq:torus_spec_general}
  \lambda_n^\pm 
  = \sqrt{
  p_\mu p_\mu + \frac{\hbar^2}{4} g^2 \vecA_\mu \vecA_\mu 
  \pm \sqrt{\hbar^2 g^2 p_\mu p_\nu \vecA_\mu \vecA_\nu 
            + \frac{\hbar^4}{4} g^4 (\vecA_1 \times \vecA_2)^2}
  } 
\end{equation}
with $p_\mu$ as in \eqref{eq:quantised_p}.
The spectrum is once more symmetric about $\lambda=0$ 
and we only show the positive eigenvalues. 

For the following we concentrate on the special case with 
\begin{equation}
\label{eq:special_color_fields}
  \vecA_1 = \begin{pmatrix} A \\ 0 \\ 0 \end{pmatrix} \, , \quad 
  \vecA_2 = \begin{pmatrix} 0 \\ A \\ 0 \end{pmatrix}
\end{equation}
in which yet none of the four contributions in 
\eqref{eq:torus_spec_general} vanishes. Introducing the abbreviation
\begin{equation}
\label{eq:definition_a}
  a := \hbar g A 
\end{equation}
we have 
\begin{equation}
  \lambda_n^\pm = 
  \sqrt{p_\mu p_\mu + \frac{a^2}{2} \pm \sqrt{a^2 p_\mu p_\mu + \frac{a^4}{4}}}
  = \sqrt{p_\mu p_\mu + \frac{a^2}{4}} \pm \frac{a}{2}
  \, .
\end{equation}
The spectral density (for positive virtuality) thus reads
\begin{equation}
\label{eq:torus_spec_density}
  \rho(\lambda) 
  = \sum_{n\in\Z^2} \left[ \delta(\lambda-\lambda_n^+) 
                           + \delta(\lambda-\lambda_n^-) \right] \, .
\end{equation}
When the spectrum is already known exactly a trace formula can usually 
be derived by employing the Poisson summation formula which 
expresses a sum over integers by a sum over the Fourier transformed addends,
\begin{equation}
  \sum_{n\in\Z^d} f(n) 
  = \sum_{k\in\Z^d} \int_{\R^d} f(n) \, \ue^{2\pi\ui k_\mu n_\mu} \, 
    \ud^d n \, .
\end{equation}
Doing this for the spectral density \eqref{eq:torus_spec_density} and 
changing variables from $n_\mu$ to 
$p_\mu = L_\mu \changed{n_\mu}/(2\pi\hbar)$ (no summation convention) we have
\begin{equation}
  \rho(\lambda) 
  = \frac{L_1 L_2}{(2\pi\hbar)^2} \sum_{k\in\Z^2} \int_{\R^2}  
  \left[ \delta(\lambda-\lambda_n^+) + \delta(\lambda-\lambda_n^-) \right]
  \ue^{\frac{\ui}{\hbar} q_\mu p_\mu} \, \ud^2p \, ,
\end{equation}
where we have introduced
\begin{equation}
  q_\mu := k_\mu L_\mu \qquad \text{(no summation convention)} \, .
\end{equation}
Introducing radial coordinates, $p=\sqrt{p_1^2+p_2^2}$, 
$q=\sqrt{k_1^2 L_1^2 + k_2^2 L_2^2}$, we obtain 
\begin{equation} 
  \rho(\lambda) 
  = \frac{L_1 L_2}{(2\pi\hbar)^2} \sum_{k\in\Z^2} \int_0^{2\pi} \int_0^\infty
    \left[ \delta(\lambda-\lambda^+(p)) + \delta(\lambda-\lambda^-(p)) \right]
    \ue^{\frac{\ui}{\hbar} q p \cos\phi} \, p \, \ud p \, \ud\phi
\end{equation}
with
\begin{equation}
  \lambda^\pm(p) := 
  \sqrt{p^2 + \frac{a^2}{4}} \pm \frac{a}{2}
\end{equation}
The $\delta$-functions select $p=\sqrt{\lambda(\lambda-a)}$ and 
$p=\sqrt{\lambda(\lambda+a)}$ in the first and second term, respectively, 
and the $\phi$-integral yields a Bessel function. Hence,
\begin{equation}
\begin{split} 
  \rho(\lambda) 
  &= \frac{L_1 L_2}{2\pi\hbar^2} 
    \left[ \Theta(\lambda-a) \left(\lambda-\frac{a}{2}\right) 
           + \left(\lambda+\frac{a}{2}\right) \right] \\[1ex]
  & \quad\,
  + \frac{L_1 L_2}{2\pi\hbar^2} \underset{k_\mu\neq0}{\sum_{k\in\Z^2}} 
  \left[ \Theta(\lambda-a) \left(\lambda-\frac{a}{2}\right) 
         J_0\left( \frac{q}{\hbar} \sqrt{\lambda(\lambda-a)} \right)
         + \left(\lambda+\frac{a}{2}\right) 
         J_0\left( \frac{q}{\hbar} \sqrt{\lambda(\lambda+a)} \right) 
  \right] \, , 
\end{split}
\end{equation}
where we have separated the mean density, 
\begin{equation}
  \overline{\rho}(\lambda) 
  = \frac{L_1 L_2}{2\pi\hbar^2} 
    \left[ \Theta(\lambda-a) \left(\lambda-\frac{a}{2}\right) 
           + \left(\lambda+\frac{a}{2}\right) \right] \, ,
\end{equation}
which derives from $k_1=k_2=0$. Notice that -- since $a$ is of order $\hbar$,
cf. \eqref{eq:definition_a} -- in leading semiclassical order the 
mean density is given by \changed{
\begin{equation}
  \overline{\rho}(\lambda) \sim \frac{L_1 L_2}{\pi\hbar^2} \, \lambda
\end{equation}
which agrees with} \eqref{eq:Weyl_hbar->0} with $d=2$, $J=2$ and,
formally, $s=0$, since there is no dynamical spin in $1+1$ dimensions.
For small $\lambda$, however, the exact mean density reads 
\begin{equation}
  \overline{\rho}(\lambda) 
  = \frac{L_1 L_2}{2\pi\hbar^2} \left(\lambda+\frac{a}{2}\right) 
  \, , \quad
  \lambda < \frac{a}{2} \, , 
\end{equation}
giving rise to a non-zero value at $\lambda=0$ which cannot be seen
by leading order semiclassical asymptotics. 

Before we can compare the periodic orbit sum with semiclassical theories 
we have to expand the result asymptotically for $\hbar\to0$. 
To this end recall that $a$ is of order $\hbar$. 
Using the asymptotic behaviour of the Bessel function we obtain 
\begin{equation}
\begin{split}
  J_0\left( \frac{q}{\hbar} \sqrt{\lambda(\lambda \pm a)} \right)
  &\sim \sqrt{\frac{2\hbar}{\pi q \sqrt{\lambda(\lambda \pm a)}}} \, 
        \cos\left( \frac{q}{\hbar} \sqrt{\lambda(\lambda+a)} 
                   - \frac{\pi}{4} \right)\\
  &\sim \sqrt{\frac{2\hbar}{\pi q \lambda}} \, 
        \cos\left( \frac{qgA}{2}\right)
        \cos\left( \frac{q}{\hbar} \lambda - \frac{\pi}{4} \right) \, ,
\end{split}
\end{equation}
and thus the semiclassical periodic orbit sum reads
\begin{equation}
\label{eq:torus_hbar->0}
  \rho_\mathrm{osc}(\lambda) 
  \sim 
  \frac{L_1 L_2}{(\pi\hbar)^{3/2}} \underset{k_\mu\neq0}{\sum_{k\in\Z^2}} 
    \sqrt{\frac{2\lambda}{q}} \, 
        \cos\left( \frac{qgA}{2}\right)
        \cos\left( \frac{q}{\hbar} \lambda - \frac{\pi}{4} \right)
\end{equation}
Let us analyse some contributions. 
The rapidly oscillating term, the last cosine, contains the argument 
$q\lambda/\hbar$. The geometric length of a periodic orbit on the torus 
with winding numbers $k_1$ and $k_2$ is given by 
$q=\sqrt{k_1^2 L_1^2 + k_2^2 L_2^2}$ and with the Hamiltonian 
\eqref{eq:free_classical_Hamiltonian} the action of this orbit reads 
\begin{equation}
  \oint p_\mu \, \ud x_\mu = \lambda q \, .
\end{equation}
The colour field shows up only in the argument of the other cosine, 
which has to derive from colour precession. 
For $\SU(2)$ fields the colour transport equation 
\eqref{eq:colour_transport} reads
\begin{equation}
  \dot{d} - \ui \frac{g p_\mu}{2\lambda} \vecA_\mu \vecsig \, d = 0 \, .
\end{equation}
With the choice \eqref{eq:special_color_fields} this becomes
\begin{equation}
  \dot{d} - \ui \frac{gA}{2\lambda} p_\mu \sigma^\mu \, d = 0 
\end{equation}
which we have to integrate with initial condition $d(0)=\eins_2$ 
up to the period of a periodic orbit on the torus. Due to the Hamiltonian
\eqref{eq:free_classical_Hamiltonian} the period equals the geometric length
of the orbit and thus the solution is given by
\begin{equation}
  d = \exp\left(-\ui \frac{gA}{2\lambda} p_\mu \sigma^\mu q\right) \, .
\end{equation}
The trace of this expression,
\begin{equation}
  \utr d 
  = 2 \cos\left( \frac{gA\sqrt{p_\mu p_\mu}}{2\lambda} q \right)
  = 2 \cos\left( \frac{gAq}{2} \right) \, , 
\end{equation} 
enters as a weight factor in the periodic orbit sum, and indeed gives rise 
to the aforesaid cosine factor. The remaining factors can be calculated 
with standard methods, see e.g. \cite{BerTab76,BerTab77a} for the general 
case or \cite[section 3.6.2]{Kep03b} for a related example.

\section{Conclusions and outlook}\label{sec:conclusions}

We have discussed the semiclassical structure of the QCD Dirac 
operator, and in particular the interplay of three semiclassical 
parameters, namely Planck's constant $\hbar$, and the spin and colour 
quantum numbers $s$ and $J$, respectively. This situation allows for 
various semiclassical scenarios, with combined semiclassical asymptotics 
considered. 
We have encountered a rich family of classical dynamics 
of translational, colour and spin degrees of freedom, 
whose mutual coupling depends on how the semiclassical limit is taken.

The influence of these different types of dynamics in semiclassical
trace formulae has been discussed and, in particular, we have analysed
the behaviour of the Weyl term, the mean density of states, in
different semiclassical scenarios. Based on this analysis we have
critically evaluated which of the semiclassical scenarios has the
potential of describing the spectrum of the QCD Dirac operator near
zero virtuality, leading us to a speculative discussion of the
mechanism \changed{behind the chiral phase transition. We certainly do
not want to overstate this speculation, a definite statement requires
further work, as indicated in section \ref{subsec:Weyl_Discussion}.}
There are various directions of research which would naturally
continue the present analysis.

So far we have mainly discussed the Weyl term, which in a trace
formula gives rise to the mean density of states, but not the periodic
orbit sum which is responsible for spectral correlations. An analysis
based on periodic orbits should, e.g., lead to a semiclassical theory
for the universal microscopic spectral density of the QCD Dirac
operator as described by chiral RMT. Moreover, such an approach would
also describe deviations from RMT behaviour on large spectral scales,
cf. saturation effects as described in \cite{Ber85}, and thus
potentially provide the missing link asked for in the introduction,
which would make it possible to directly use RMT information when
calculating fermionic determinants. Here, one should keep in mind that
the equivalent of the Thouless energy~\cite{JanNowPapZah98,OsbVer98}
sets another scale.  It is not present in Dirac spectra for frozen
gauge fields, but it is an ensemble effect resulting from the
propagation of the gauge fields.

Besides the various semiclassical scenarios which we have described
in this article there is an additional strategy for taking the 
semiclassical limit of multi-component wave equations. In this 
approach one does not treat the matrix degrees of freedom, 
i.e. colour and spin, dynamically but rather considers polarised 
Hamiltonians, which describe a particle with the spin or colour projection 
locked to the ``direction'' of the external gauge field, 
see  e.g. \cite{LitFly91b,FriGuh93,BolKep99a,AmaBra02}. Such an approach
may also prove useful in the case of the QCD Dirac operator. 

In lattice gauge theory, which we have referred to in various places,  
the Dirac operator is implemented as a difference instead of a differential 
operator. This has consequences which could also be analysed within the 
semiclassical picture. On the one hand discretisation leads to a modified 
dispersion relation, i.e. to different classical Hamiltonians $\Lambda$.
Roughly speaking, the momenta are replaced by suitably normalised sines 
of momenta, which in the semiclassical picture changes both, Weyl terms and 
the periodic orbit structure. On the other hand the discretised theory lives
in a finite dimensional Hilbert space. In a semiclassical context the 
dimension of this Hilbert space also becomes a semiclassical parameter
(cf. the theory of quantised maps, \cite{HanBer80}, see also \cite{DeB01} 
for an overview), which would allow for the continuum limit to be discussed 
on a semiclassical footing. 

\vspace*{0.5cm}
\subsection*{Acknowledgement}
We thank Tilo Wettig for numerous stimulating discussions and helpful
remarks. Moreover we benefited from useful discussions with Johan
Bijnens, Jens Bolte, Dmitri Diakonov, Stephen Fulling, and Ed Shuryak.
TG acknowledges support from Det Svenska Vetenskapsr{\aa}det 
and SK is grateful for support from Deutsche Forschungsgemeinschaft
under grant no.~KE~888/1-1 and also from Crafoordska Stiftelsen under
grant no.~20020681.

\begin{appendix}
\section{Projected transport equations}
\label{app:transport}

In order to calculate the projected transport transport equations 
\eqref{eq:projected_transport} for the $2\times2$ matrices $b_\pm$ 
we have to evaluate the expressions,  
\begin{equation}
  V_\pm^\dag \gamma_\mu V_\pm (\partial_\mu b_\pm) 
  \, , \quad
  V_\pm^\dag V_\pm \frac{\partial b_\pm}{\partial t} 
  \, , \quad
  V_\pm^\dag \frac{\partial}{\partial t} V_\pm
  \quad \text{and} \quad
  V_\pm^\dag \gamma_\mu \partial_\mu V_\pm^\dag \, .
\end{equation}
We begin with the terms where the derivatives act on 
$b_\pm$: 
\begin{equation}
\begin{split}
  V_+^\dag \gamma_\mu a_\mu V_+ 
  &= \frac{1}{2} 
     \left( \eins_2 \, , \ \tfrac{\pi_4-\ui\vecsig\vecpi}{\Lambda} \right) 
     \begin{pmatrix} 0 & a_4-\ui\vecsig\veca \\ 
                     a_4+\ui\vecsig\veca & 0 \end{pmatrix}
     \begin{pmatrix} \eins_2 \\ \frac{\pi_4 + \ui \vecsig\vecpi}{\Lambda} 
     \end{pmatrix}\\
  &= \frac{1}{2} 
     \left[ \frac{\pi_4-\ui\vecsig\vecpi}{\Lambda} (a_0+\ui\vecsig\veca) 
     + (a_0-\ui\vecsig\veca) \frac{\pi_4+\ui\vecsig\vecpi}{\Lambda} \right]\\
  &= \frac{1}{2\Lambda} ( 2a_0\pi_4 + 2 \veca\vecpi) 
   = \frac{\pi_\mu a_\mu}{\Lambda}\\
  \Longrightarrow \quad & V_+^\dag \gamma_\mu V_+ (\partial_\mu b_+) 
  = \frac{\partial \Lambda}{\partial p_\mu} (\partial_\mu b_+) \, .
\end{split}
\end{equation}
Due to normalisation, $V_\pm^\dag V_\pm = \eins_2$, 
the second contribution is trivial, and together with the first one yields 
\begin{equation}
  V_+^\dag V_+ \frac{\partial b_+}{\partial t} 
  + V_+^\dag \gamma_\mu V_+ (\partial_\mu b_+) 
  = \left( \frac{\partial}{\partial t} 
    + \frac{\pi_\mu}{\Lambda} \partial_\mu \right) b_+ 
  =: \dot{b}_+ \, , 
\end{equation} 
where the dot denotes a derivative along the flow with Hamiltonian $\Lambda$.
The remaining two terms yield
\begin{equation}
\label{eq:V_+^+d/dtV_+}
\begin{split}
  V_+^\dag \frac{\partial}{\partial t} V_+ 
  &= V_+^\dag \frac{1}{\sqrt{2}} \begin{pmatrix} 0 \\
     \frac{\frac{\partial \pi_4}{\partial t} 
           + \ui \vecsig \frac{\partial \vecpi}{\partial t}}{\Lambda} 
     - \frac{\pi_4+\ui\vecsig\vecpi}{\Lambda^3} \pi_\mu 
       \frac{\partial \pi_\mu}{\partial t} \end{pmatrix}\\
  &= \frac{1}{2} \left[ 
     \frac{1}{\Lambda^2} (\pi_4 -\ui \vecsig \vecpi )
     \left( \frac{\partial \pi_4}{\partial t} 
            + \ui \vecsig \frac{\partial \vecpi}{\partial t} \right) 
     - \frac{1}{\Lambda^2} \pi_\mu \frac{\partial \pi_\mu}{\partial t} 
     \right]\\
  &= \frac{1}{2\Lambda^2} \left[ 
     \ui \pi_4 \vecsig\frac{\partial \vecpi}{\partial t} 
     - \ui \vecsig\vecpi \frac{\partial \pi_4}{\partial t} 
     + \ui \vecsig \left( \vecpi \times \frac{\vecpi}{\partial t} \right) 
     \right] 
\end{split}
\end{equation}
and  
\begin{equation}
\begin{split}
  V_+^\dag \gamma_\mu \partial_\mu V_+^\dag
  &= V_+^\dag \frac{1}{\sqrt{2}} \begin{pmatrix}
     (\partial_0 - \ui \vecsig \nabla) 
     \frac{\pi_4+\ui\vecsig\vecpi}{\Lambda} \\ 0  \end{pmatrix}\\
  &= \frac{1}{2} \left[ (\partial_0 - \ui \vecsig \nabla) 
     \frac{\pi_4+\ui\vecsig\vecpi}{\Lambda} \right]\\
  &= \frac{1}{2} \left[ 
     - \frac{1}{\Lambda^3} \pi_\mu (\partial_0\pi_\mu-\ui\vecsig\nabla\pi_\mu)
       (\pi_4+\ui\vecsig\vecpi) 
     + \frac{1}{\Lambda} (\partial_0 - \ui \vecsig \nabla) 
       (\pi_4+\ui\vecsig\vecpi) \right] \, .
\end{split}
\end{equation}
Using the Hamilton-Jacobi equation \eqref{eq:HJG_abelian} we can derive the 
useful relation 
\begin{equation}
  \frac{\partial\pi_\mu}{\partial t} 
  = \frac{\partial}{\partial t} \partial_\mu S 
  = \partial_\mu \frac{\partial S}{\partial t} 
  = -\partial_\mu \Lambda 
  = - \frac{\pi_\nu}{\Lambda} \partial_\mu \pi_\nu \, , 
\end{equation} 
which we now use ``backwards'', 
\begin{equation}
  - \frac{1}{\Lambda^3} \pi_\mu (\partial_0\pi_\mu-\ui\vecsig\nabla\pi_\mu)
  = \frac{1}{\Lambda^2} \left( \frac{\partial\pi_4}{\partial t} 
    - \ui\vecsig \frac{\partial\vecpi}{\partial t} \right) \, . 
\end{equation}
Hence, 
\begin{equation}
\begin{split}
  V_+^\dag \gamma_\mu \partial_\mu V_+^\dag
  &= \frac{1}{2} \left[ \frac{1}{\Lambda^2} \left( 
     \pi_4\frac{\partial\pi_4}{\partial t} 
     + \ui \vecsig\vecpi \frac{\partial\pi_4}{\partial t}
     - \ui \pi_4 \vecsig\frac{\partial\vecpi}{\partial t}
     + \vecpi \frac{\partial\vecpi}{\partial t}
     + \ui\vecsig\left(\frac{\partial\vecpi}{\partial t}\times\vecpi\right)
     \right) \right. \\ & \left. \qquad \qquad 
     + \frac {1}{\Lambda} \left( 
     \partial_0 \pi_4 + \ui \vecsig \partial_0 \vecpi - \ui \vecsig \nabla \pi_4 + \nabla \vecpi + \ui \vecsig (\nabla \times \vecpi) \right) \right] \, , 
\end{split}
\end{equation}
and added to \eqref{eq:V_+^+d/dtV_+} a couple of terms drop out,
\begin{equation}
\begin{split}
  V_+^\dag \frac{\partial}{\partial t} V_+ 
  + V_+^\dag \gamma_\mu \partial_\mu V_+^\dag
  &= \underbrace{
     \frac{\pi_\mu}{2\Lambda^2} \frac{\partial \pi_\mu}{\partial t} 
     + \frac{\partial_\mu \pi_\mu}{2\Lambda}}_{%
       \frac{1}{2}\left[\frac{\partial^2\Lambda}{\partial x_\mu \partial p_\mu}
       + \frac{\partial^2\Lambda}{\partial p_\mu \partial p_\nu}
         \frac{\partial^2 S}{\partial x_\mu \partial x_\nu}\right]}
     + \frac{\ui \vecsig}{2\Lambda} 
       (\partial_0 \vecpi - \nabla \pi_4 + \nabla \times \vecpi) \, .
\end{split}
\end{equation}
With 
\begin{equation}
\label{eq:E-field}
  \partial_0 \vecpi - \nabla \pi_4 
  = \partial_0 (\nabla S - g \vecA) - \nabla (\partial_0 S - g A_0)
  = -g ( \partial_0 \vecA - \nabla A_0) 
  = -g \vecE
\end{equation}
and 
\begin{equation}
\label{eq:B-field}
  \nabla \times \vecpi 
  = \nabla \times (\nabla S - g \vecA) 
  = -g \nabla \times \vecA 
  = -g \vecB
\end{equation}
the non-scalar terms finally are given by
\begin{equation}
  -\frac{\ui g}{2\Lambda} \vecsig (\vecB + \vecE) \, .
\end{equation}

The analogous calculation for $b_-$ reads
\begin{equation}
\begin{split}
  V_-^\dag \gamma_\mu a_\mu V_- 
  &= \frac{1}{2} 
     \left( \tfrac{\pi_4+\ui\vecsig\vecpi}{\Lambda} \, , \ -\eins_2 \right) 
     \begin{pmatrix} 0 & a_0-\ui\vecsig\veca \\ 
                     a_0+\ui\vecsig\veca & 0 \end{pmatrix}
     \begin{pmatrix} \frac{\pi_4 - \ui \vecsig\vecpi}{\Lambda} \\ -\eins_2
     \end{pmatrix}\\
  &= \frac{1}{2} \left[
     - \frac{\pi_4+\ui\vecsig\vecpi}{\Lambda} (a_0-\ui\vecsig\veca) 
     - (a_0+\ui\vecsig\veca) \frac{\pi_4-\ui\vecsig\vecpi}{\Lambda} \right]\\
  &= -\frac{1}{2\Lambda} ( 2a_0\pi_4 + 2 \veca\vecpi) 
   = -\frac{\pi_\mu a_\mu}{\Lambda}\\
  \Longrightarrow \quad & V_-^\dag \gamma_\mu V_- (\partial_\mu b_-) 
  = \frac{\partial \Lambda^-}{\partial p_\mu} (\partial_\mu b_-) \\
  \Longrightarrow \quad &
  V_-^\dag V_- \frac{\partial b_-}{\partial t} 
  + V_-^\dag \gamma_\mu V_- (\partial_\mu b_-) 
  = \left( \frac{\partial}{\partial t} 
    + \frac{\pi_\mu}{\Lambda^-} \partial_\mu \right) b_- 
  =: \dot{b}_- \, , 
\end{split}
\end{equation} 
where the dot now denotes a derivative along the flow with Hamiltonian 
$\Lambda^-$. The other two terms yield
\begin{equation}
\label{eq:V_-^+d/dtV_-}
\begin{split}
  V_-^\dag \frac{\partial}{\partial t} V_- 
  &= V_-^\dag \frac{1}{\sqrt{2}} \begin{pmatrix} 
     \frac{\frac{\partial \pi_4}{\partial t} 
     - \ui \vecsig \frac{\partial \vecpi}{\partial t}}{\Lambda} 
     - \frac{\pi_4-\ui\vecsig\vecpi}{\Lambda^3} \pi_\mu 
       \frac{\partial \pi_\mu}{\partial t} \end{pmatrix}\\
  &= \frac{1}{2} \left[ 
     \frac{1}{\Lambda^2} (\pi_4+\ui \vecsig \vecpi )
     \left( \frac{\partial \pi_4}{\partial t} 
     - \ui \vecsig \frac{\partial \vecpi}{\partial t} \right) 
     - \frac{1}{\Lambda^2} \pi_\mu \frac{\partial \pi_\mu}{\partial t} 
     \right]\\
  &= \frac{1}{2\Lambda^2} \left[ 
     - \ui \pi_4 \vecsig\frac{\partial \vecpi}{\partial t} 
     + \ui \vecsig\vecpi \frac{\partial \pi_4}{\partial t} 
     + \ui \vecsig \left( \vecpi \times \frac{\vecpi}{\partial t} \right) 
     \right] 
\end{split}
\end{equation}
and 
\begin{equation}
\begin{split}
  V_-^\dag \gamma_\mu \partial_\mu V_-^\dag
  &= V_-^\dag \frac{1}{\sqrt{2}} \begin{pmatrix} 0\\
     (\partial_0 + \ui \vecsig \nabla) 
     \frac{\pi_4-\ui\vecsig\vecpi}{\Lambda} \\ 0  \end{pmatrix}\\
  &= - \frac{1}{2} \left[ (\partial_0 + \ui \vecsig \nabla) 
     \frac{\pi_4-\ui\vecsig\vecpi}{\Lambda} \right]\\
  &= -\frac{1}{2} \left[ 
     - \frac{1}{\Lambda^3} \pi_\mu (\partial_0\pi_\mu+\ui\vecsig\nabla\pi_\mu)
       (\pi_4-\ui\vecsig\vecpi) 
     + \frac{1}{\Lambda} (\partial_0 + \ui \vecsig \nabla) 
       (\pi_4-\ui\vecsig\vecpi) \right] \, .
\end{split}
\end{equation}
Again we use the Hamilton-Jacobi equation \eqref{eq:HJG_abelian},
\begin{equation}
  \frac{\partial\pi_\mu}{\partial t} 
  = \frac{\partial}{\partial t} \partial_\mu S 
  = \partial_\mu \frac{\partial S}{\partial t} 
  = -\partial_\mu \Lambda^-
  = \partial_\mu \Lambda 
  = \frac{\pi_\nu}{\Lambda} \partial_\mu \pi_\nu \, , 
\end{equation} 
concluding that 
\begin{equation}
\begin{split}
  V_-^\dag \gamma_\mu \partial_\mu V_-^\dag
  &= -\frac{1}{2} \left[ -\frac{1}{\Lambda^2} \left( 
     \pi_4\frac{\partial\pi_4}{\partial t} 
     - \ui \vecsig\vecpi \frac{\partial\pi_4}{\partial t}
     + \ui \pi_4 \vecsig\frac{\partial\vecpi}{\partial t}
     + \vecpi \frac{\partial\vecpi}{\partial t}
     + \ui\vecsig\left(\frac{\partial\vecpi}{\partial t}\times\vecpi\right)
     \right) \right. \\ & \left. \qquad \qquad 
     + \frac {1}{\Lambda} \left( 
     \partial_0 \pi_4 - \ui \vecsig \partial_0 \vecpi 
     + \ui \vecsig \nabla \pi_4 + \nabla \vecpi 
     + \ui \vecsig (\nabla \times \vecpi) \right) \right] \, , 
\end{split}
\end{equation}
and together with \eqref{eq:V_-^+d/dtV_-} we obtain
\begin{equation}
\begin{split}
  V_-^\dag \frac{\partial}{\partial t} V_- 
  + V_-^\dag \gamma_\mu \partial_\mu V_-^\dag
  &= \underbrace{
     \frac{\pi_\mu}{2\Lambda^2} \frac{\partial \pi_\mu}{\partial t} 
     - \frac{\partial_\mu \pi_\mu}{2\Lambda}}_{%
       \frac{1}{2}\left[\frac{\partial^2\Lambda^-}
                             {\partial x_\mu \partial p_\mu}
       + \frac{\partial^2\Lambda^-}{\partial p_\mu \partial p_\nu}
         \frac{\partial^2 S}{\partial x_\mu \partial x_\nu}\right]}
     + \frac{\ui \vecsig}{2\Lambda} 
       (\partial_0 \vecpi - \nabla \pi_4 - \nabla \times \vecpi) \, .
\end{split}
\end{equation}
With \eqref{eq:E-field} and \eqref{eq:B-field} the non-scalar terms
in this case read
\begin{equation}
 -\frac{\ui g}{2\Lambda} \vecsig(\vecB-\vecE)
 = -\frac{\ui g}{2\Lambda^-} \vecsig(\vecE-\vecB) \, .
\end{equation}

\changed{
\section{Local Gaussian fluctuations vs.~constant random fields}
\label{app:random_measures}
We show that averaging Weyl terms over stochastic fields -- more
precisely, independent locally Gaussian fields -- is computationally
equivalent to averaging over constant Gaussian fields.

When averaging a mean density $\overline{\rho}$, given by a Weyl term,
over fields, say $B(x)$, it is crucial that the Weyl term is given by
an integral over the position variable $x$, see
\eqref{eq:Weyl_general}. 
Therefore, let us now consider expressions of the form
\begin{equation}
  I := \int \int f(B(x),x) \, \ud x \, \mathcal{D}B
\end{equation}
with a Gaussian measure $\mathcal{D}B$. Think of the functional
integral as defined by a suitably normalised continuum limit  
$N \to \infty$ of its discretised analogue, 
\begin{equation}
  I_N := \int\cdots\int \sum_{j=1}^N f(B_j,x_j) \, \prod_{k=1}^N
  \frac{\ue^{-B_k^2/(2\sigma^2)}}{\sqrt{2\pi\sigma^2}} \, \ud B_k \, ,
\end{equation}
with lattice points $x_j$ and $B(x_j)=:B_j$. Then we obtain
\begin{equation}
  I_N  
  = \sum_{j=1}^N \int f(B_j,x_j) \, 
  \frac{\ue^{-B_j^2/(2\sigma^2)}}{\sqrt{2\pi\sigma^2}} \, \ud B_j 
  = \int \sum_{j=1}^N  f(B,x_j) \, 
  \frac{\ue^{-B^2/(2\sigma^2)}}{\sqrt{2\pi\sigma^2}} \, \ud B \, . 
\end{equation}
In the continuum limit we have thus derived the relation
\begin{equation}
  \int \int f(B(x),x) \, \ud x \, \mathcal{D}B 
  = \int \int f(B,x) \,
  \frac{\ue^{-B^2/(2\sigma^2)}}{\sqrt{2\pi\sigma^2}} \, 
  \ud x \, \ud B \, , 
\end{equation}
i.e. when interested in local Gaussian fluctuations in section
\ref{subsec:Weyl_triple_limit}, we may average over constant random
fields instead.

}

\end{appendix}

\bibliographystyle{my_unsrt}                   
\bibliography{literatur}                       

\begin{thebibliography}{10}

\bibitem{ShuVer93}
E.~V. Shuryak and J.~J.~M. Verbaarschot: {\em Random matrix theory and spectral
  sum rules for the {D}irac operator in {QCD}\/}, Nucl. Phys. A {\bf 560}
  (1993) ~306--320.

\bibitem{VerZah93}
J.~J.~M. Verbaarschot and I.~Zahed: {\em Spectral density of the QCD Dirac
  operator near zero virtuality\/}, Phys. Rev. Lett. {\bf 70} (1993)
  ~3852--3855.

\bibitem{VerWet00}
J.~J.~M. Verbaarschot and T.~Wettig: {\em Random Matrix Theory and Chiral
  Symmetry in QCD\/}, Ann. Rev. Nucl. Part. Sci. {\bf 50} (2000) ~343--410.

\bibitem{JanNowPapZah98}
R.~A. Janik, M.~A. Nowak, G.~Papp and I.~Zahed: {\em Chiral {D}isorder in
  {QCD}\/}, Phys. Rev. Lett. {\bf 81} (1998) ~264--267.

\bibitem{OsbVer98}
J.~C. Osborn and J.~J.~M. Verbaarschot: {\em Thouless Energy and Correlations
  of QCD Dirac Eigenvalues\/}, Phys. Rev. Lett. {\bf 81} (1998) ~268--271.

\bibitem{BerGoeGuhJacMaMeySchWeiWetWil98}
M.~Berbenni-Bitsch, M.~G\"ockeler, T.~Guhr, A.~Jackson, J.~Ma, S.~Meyer,
  A.~Sch\"afer, H.~Weidenm\"uller, T.~Wettig and T.~Wilke: {\em The Range of
  Validity for the Random Matrix Description of Lattice Gauge Theories in the
  Microscopic Regime\/}, Phys. Lett. B {\bf 438} (1998) ~14--20.

\bibitem{GuhMaMeyWil99}
T.~Guhr, J.-Z. Ma, S.~Meyer and T.~Wilke: {\em Statistical analysis and the
  equivalent of a {T}houless energy in lattice QCD Dirac spectra\/}, Phys. Rev.
  D {\bf 59} (1999) ~054501.

\bibitem{Gut71}
M.~C. {G}utzwiller: {\em {P}eriodic {O}rbits and {C}lassical {Q}uantization
  {C}onditions\/}, J. Math. Phys. {\bf 12} (1971) ~343--358.

\bibitem{Ber85}
M.~V. {B}erry: {\em {S}emiclassical theory of spectral rigidity\/}, Proc. R.
  Soc. London Ser. A {\bf 400} (1985) ~229--251.

\bibitem{GuhWilWei00}
T.~Guhr, T.~Wilke and H.~A. Weidenm\"uller: {\em Stochastic Field Theory for a
  Dirac Particle Propagating in Gauge Field Disorder\/}, Phys. Rev. Lett. {\bf
  85} (2000) ~2252--2255.

\bibitem{GuhWil01b}
T.~Guhr and T.~Wilke: {\em Non-linear $\sigma$ model for gauge field
  disorder\/}, Nucl. Phys. B {\bf 593} (2001) ~361--397.

\bibitem{Kep03b}
S.~Keppeler: {\em Spinning Particles: Semiclassics and Spectral Statistics\/},
  no. 193 in Springer Tracts in Modern Physics, Springer-Verlag, Berlin
  Heidelberg,  (2003).

\bibitem{Mei92}
E.~Meinrenken: {\em {S}emiclassical principal symbols and {G}utzwiller's trace
  formula\/}, Rep. Math. Phys. {\bf 31} (1992) ~279--295.

\bibitem{PauUri95}
T.~Paul and A.~Uribe: {\em The Semi-Classical Trace Formula and Propagation of
  Wave Packets\/}, J. Funct. Anal. {\bf 132} (1995) ~192--249.

\bibitem{BolKep99a}
J.~{B}olte and S.~{K}eppeler: {\em A semiclassical approach to the {D}irac
  equation\/}, Ann. Phys. (NY) {\bf 274} (1999) ~125--162.

\bibitem{BolKep98}
J.~{B}olte and S.~{K}eppeler: {\em {S}emiclassical Time Evolution and Trace
  Formula for Relativistic Spin-1/2 Particles\/}, Phys. Rev. Lett. {\bf 81}
  (1998) ~1987--1991.

\bibitem{Tho27}
L.~H. Thomas: {\em The Kinematics of an Electron with an Axis\/}, Philos. Mag.
  {\bf 3} (1927) ~1--22.

\bibitem{BarMicTel59}
V.~Bargman, L.~Michel and V.~L. Telegdi: {\em Precession of the polarization of
  particles moving in a homogeneous electromagnetic field\/}, Phys. Rev. Lett.
  {\bf 2} (1959) ~435--436.

\bibitem{Kir76}
A.~A. Kirillov: {\em Elements of the theory of representations\/}, no. 220 in
  Grundlehren der Mathematischen Wissenschaften, Springer-Verlag, Berlin,
  (1976).

\bibitem{BolGla04}
J.~Bolte and R.~Glaser: {\em A semiclassical {E}gorov theorem and quantum
  ergodicity for matrix valued operators\/}, Comm. Math. Phys. {\bf 247} (2004)
  ~391--419.

\bibitem{BolKep99b}
J.~{B}olte and S.~{K}eppeler: {\em {S}emiclassical form factor for chaotic
  systems with spin 1/2\/}, J. Phys. A {\bf 32} (1999) ~8863--8880.

\bibitem{BolGlaKep01}
J.~Bolte, R.~Glaser and S.~Keppeler: {\em Quantum and classical ergodicity of
  spinning particles\/}, Ann. Phys. (NY) {\bf 293} (2001) ~1--14.

\bibitem{Haa01}
F.~Haake: {\em {Q}uantum {S}ignatures of {C}haos\/}, {S}pringer-{V}erlag,
  {B}erlin {H}eidelberg, 2nd edn.,  (2001).

\bibitem{PleAmaMehBra02}
M.~Pletyukhov, C.~Amann, M.~Mehta and M.~Brack: {\em Semiclassical theory of
  spin-orbit interactions using spin coherent states\/}, Phys. Rev. Lett. {\bf
  89} (2002) ~116601.

\bibitem{PleZai03}
M.~Pletyukhov and O.~Zaitsev: {\em Semiclassical theory of spin-orbit
  interaction in the extended phase space\/}, J. Phys. A {\bf 36} (2003)
  ~5181--5210.

\bibitem{BolGla05}
J.~Bolte and R.~Glaser: {\em Semiclassical propagation of coherent states with
  spin-orbit interaction\/}, Ann. H. Poincar\'e {\bf 6} (2005) ~625--656.

\bibitem{tHo74}
G.~'t~Hooft: {\em A planar diagram theory for strong interactions\/}, Nucl.
  Phys. B {\bf 72} (1974) ~461--473.

\bibitem{Won70}
S.~K. Wong: {\em Field and particle equations for the classical {Y}ang-{M}ills
  field and particles with isotopic spin\/}, Nuovo Cimento A {\bf 65} (1970)
  ~689--694.

\bibitem{BoyPerSan01}
L.~J. Boya, A.~M. Perelomov and M.~Santander: {\em Berry phase in homogeneous
  {K}\"ahler manifolds with linear {H}amiltonians\/}, J. Math. Phys. {\bf 42}
  (2001) ~5130--5142.

\bibitem{Mon84}
R.~Montgomery: {\em Canonical formulations of a classical particle in a
  {Y}ang-{M}ills field and {W}ong's equations\/}, Lett. Math. Phys. {\bf 8}
  (1984) ~59--67.

\bibitem{BanCas80}
T.~Banks and A.~Casher: {\em Chiral symmetry breaking in confining theories\/},
  Nucl. Phys. B {\bf 169} (1980) ~103--125.

\bibitem{SmiSte93}
A.~V. Smilga and J.~Stern: {\em On the spectral density of Euclidean Dirac
  operator in QCD\/}, Phys. Lett. B {\bf 318} (1993) ~531--536.

\bibitem{Zya00}
K.~Zyablyuk: {\em {D}irac operator spectral density and low energy sum
  rules\/}, J. High Energy Phys. {\bf 06} (2000) ~25.

\bibitem{ShuSmi97}
I.~A. Shushpanov and A.~V. Smilga: {\em Quark condensate in a magnetic
  field\/}, Phys. Lett. B {\bf 402} (1997) ~351--358.

\bibitem{MonMue97}
I.~Montvay and G.~M\"unster: {\em Quantum Fields on a Lattice\/}, Cambridge
  University Press, Cambridge,  (1997).

\bibitem{LeuSmi92}
H.~Leutwyler and A.~Smilga: {\em Spectrum of {D}irac operator and role of
  winding number in {QCD}\/}, Phys. Rev. D {\bf 46} (1992) ~5607--5632.

\bibitem{GnuSeiOppZir03}
S.~Gnutzmann, B.~Seif, F.~{von Oppen} and M.~R. Zirnbauer: {\em Universal
  spectral statistics of {A}ndreev billiards: {S}emiclassical approach\/},
  Phys. Rev. E {\bf 67} (2003) ~046225.

\bibitem{GnuSei04a}
S.~Gnutzmann and B.~Seif: {\em Universal spectral statistics in
  {W}igner-{D}yson, chiral, and {A}ndreev star graphs. {I}. {C}onstruction and
  numerical results\/}, Phys. Rev. E {\bf 69} (2004) ~056219.

\bibitem{GnuSei04b}
S.~Gnutzmann and B.~Seif: {\em Universal spectral statistics in
  {W}igner-{D}yson, chiral, and {A}ndreev star graphs. {II}. {S}emiclassical
  approach\/}, Phys. Rev. E {\bf 69} (2004) ~056220.

\bibitem{BerTab76}
M.~V. {B}erry and M.~{T}abor: {\em {C}losed orbits and the regular bound
  spectrum\/}, Proc. R. Soc. London Ser. A {\bf 349} (1976) ~101--123.

\bibitem{BerTab77a}
M.~V. {B}erry and M.~{T}abor: {\em {C}alculating the bound spectrum by path
  summation in action-angle variables\/}, J. Phys. A {\bf 10} (1977) ~371--379.

\bibitem{LitFly91b}
R.~G. Littlejohn and W.~G. Flynn: {\em {G}eometric phases in the asymptotic
  theory of coupled wave equations\/}, Phys. Rev. A {\bf 44} (1991)
  ~5239--5256.

\bibitem{FriGuh93}
H.~Frisk and T.~Guhr: {\em Spin-Orbit Coupling in Semiclassical
  Approximation\/}, Ann. Phys. (NY) {\bf 221} (1993) ~229--257.

\bibitem{AmaBra02}
C.~Amann and M.~Brack: {\em Semiclassical trace formulae for systems with
  spin-orbit interactions: successes and limitations of present approaches\/},
  J. Phys. A {\bf 35} (2002) ~6009--6032.

\bibitem{HanBer80}
J.~H. {H}annay and M.~V. {B}erry: {\em Quantization of linear maps on a torus
  -- Fresnel Diffraction by a periodic grating\/}, Physica D {\bf 1} (1980)
  ~267--290.

\bibitem{DeB01}
S.~De{B}ievre: {\em Quantum chaos: a brief first visit\/}, in: {\em Second
  Summer School in Analysis and Mathematical Physics: Topics in Analysis:
  Harmonic, Complex, Nonlinear and Quantization\/} (Eds. S.~Perez-Esteva and
  C.~Villegas-Blas), vol. 289 of {\em Contemp. Math.\/},  161--218, Providence,
   (2001), American Mathematical Society, mp\_arc 01-207.

\end{thebibliography}

\end{document}